\documentclass[10pt,final]{iopart}
\usepackage[dvips]{graphicx}
\usepackage{subeqn}
\newcommand{\bm}[1]{\mathbf{#1}} 
\newcommand{\pp}[2]{\frac{\partial #1}{\partial #2}}
\newcommand{\eq}[1]{(\ref{#1})}
\newcommand{\mM}{\mathit{M}} 
\newcommand{\TrM}{\Tr\mM}
\renewcommand{\Re}{\mathop{\rm Re}}

\newcommand{\sgn}{\mathop{\rm sgn}}
\newcommand{\NN}{\nonumber\\}
\newcommand{\cE}{{\mathcal{E}}}
\newcommand{\cK}{{\mathcal{K}}}
\newcommand{\rmC}{\mathrm{C}}
\newcommand{\rmS}{\mathrm{S}}
\eqnobysec

\begin{document}

\title{Normal forms and uniform approximations
  for bridge orbit bifurcations}
       
\author[K. Arita and M. Brack]{%
  Ken-ichiro Arita$^{1,2}$  and Matthias Brack$^2$}

\address{$^1$ Department of Physics,
  Nagoya Institute of Technology, \\ 466-8555 Nagoya, Japan}
\address{$^2$ Institute for Theoretical Physics,
  University of Regensburg, \\ D-93040 Regensburg, Germany}

\eads{\mailto{arita@nitech.ac.jp},
      \mailto{Matthias.Brack@physik.uni-regensburg.de}}

\begin{abstract}
We discuss various bifurcation problems in which two isolated periodic
orbits exchange periodic ``bridge'' orbit(s) between two successive
bifurcations.  We propose normal forms which locally describe the
corresponding fixed point scenarios on the Poincar\'e surface of
section.  Uniform approximations for the density of states for an
integrable Hamiltonian system with two degrees of freedom are derived
and successfully reproduce the numerical quantum-mechanical results.
\end{abstract}
\pacs{05.45.Mt, 03.65.Sq}

\section{Introduction}

The periodic orbit theory \cite{Gutzwiller,BalBlo,POT} has made
important contributions to the under\-standing of quantum chaos
\cite{Gutzbook,Stockmann,Haake} and to the semiclassical
interpretation of quantum shell effects in finite fermion systems
\cite{BrackText}.  Through semiclassical trace formulae, it relates 
the density of states of a quantum Hamiltonian to the sum over all 
periodic orbits of the corresponding classical Hamiltonian system. 
Gutzwiller's trace formula \cite{Gutzwiller} assumes the periodic 
orbits to be isolated and therefore applies most directly to 
chaotic systems; it can, however, also be used for integrable 
systems with isolated orbits (see e.g., \cite{BraJain,Sieber97}). 

In the derivation of the trace formula \cite{Gutzwiller}, the
stationary phase approximation is used for evaluating some of the
trace integrals over the semiclassical amplitude, lead\-ing to
Gauss-Fresnel integrals.  In systems with regular or mixed classical
dynamics, periodic orbits can undergo bifurcations at critical values
of a system parameter (e.g., energy or deformation).  At such critical
points, one or more of the Gauss-Fresnel integrals become singular and
cause the divergence of the Gutzwiller amplitudes of the bifurcating
orbits.  This situation can be remedied \cite{Ozorio87} by going to
higher than second-order terms in the expansion of the action
function, which appears in the phase of the trace integrand, around
the critical point in phase space.  The minimum number of terms
required in this expansion are given by the so-called normal forms,
which are characteristic for each type of bifurcation (see also
\cite{Ozobook}) and lead to usually well-known catastrophe integrals.
At the bifurcation points, one obtains in this way {\it local
uniform approximations} for the semiclassical amplitudes that are
finite and contain the contributions of all orbits involved in the
bifurcation. 

The local uniform approximations do, however, not reproduce the
correct asymp\-totic Gutzwiller amplitudes far away from the
bifurcation point where all involved orbits are isolated.  To achieve
this goal, {\it global uniform approximations} must be devel\-oped
which interpolate smoothly from the local behavior at the bifurcation
to the asymptotic regions of the isolated orbits.  This has been done
in Refs.~\cite{Sieber96,Schomerus97a,Sieber98} for all generic
bifurcations occurring in Hamiltonian systems with two degrees of
freedom according to the classification of Meyer \cite{Meyer} (and
listed also in \cite{Ozobook}), in Refs.~\cite{Schomerus97b,Kaidel04a}
for codimension-two bifurcations, and in Ref.~\cite{Transbif} for the
transcritical bifurcation.  (In passing, we mention that similar
divergences of the Gutzwiller amplitudes occur when a symmetry is
broken -- or restored -- under the variation of a system parameter.
Local uniform approximations for symmetry breaking have been developed
in Refs.~\cite{Ozorio87,Creagh96}; the prototype of a global uniform
approximation for the breaking of U(1) symmetry was developed in
Ref.~\cite{Tomsovic}, which inspired those mentioned above for
bifurcations as well as global uniform approximations for the breaking
of other symmetries \cite{Braunifo}.)

In this paper we investigate a type of bifurcation that has not yet
been studied in this context and that we term {\it bridge orbit
bifurcation}.  Typically, this bifurcation consists of a pair of
isolated orbits which are connected through a ``bridge orbit'' that
only exists in a finite interval of the system parameter.  Under its
monotonous variation, the bridge orbit is born at a bifurcation of the
first isolated orbit and then absorbed at a bifurcation of the second
isolated orbit.  This scenario has been found in both integrable and
non-integrable Hamiltonian systems\cite{Arita04,Arita06,ErikDahl}.
It occurs, e.g., in the
two-dimensional rationally deformed harmonic oscillator under a
generic class of perturbations; other examples will be given in
section \ref{sec:bridge}.  In the integrable case the bridge orbit
forms a continuously degenerate family (i.e., a rational torus), while
in non-integrable cases it is typically isolated.  Since the two
isolated orbits which exchange the bridge orbit typically are well
separated in phase space, the bridge bifurcation is accompanied by
global changes of the phase space structure which cannot be treated
with the usual perturbative normal forms derived from the
Birkoff-Gustavson expansion.  This is different from the generic
bifurcations of codimension one \cite{Ozobook}, and also from those of
codimension two considered in \cite{Schomerus97b,Kaidel04a}, where
only local changes in the phase space structure occur around a central
periodic orbit that exists at all values of the relevant system
parameter.

In all the bifurcation types investigated so far in connection with
uniform approximations, {\it all} orbits participating in the
bifurcations become asymptotically isolated far enough from the
bifurcation point(s).  Even in the codimension-two bifurcations
considered in \cite{Schomerus97b,Kaidel04a}, where some isolated
orbits only exist between two adjacent bifurcations, these have their
well-defined partners which become asymptotically isolated at least on
one side of one of the bifurcations.  For the bridge orbits studied
here, this is not the case.  The bridge orbits are entirely intrinsic
to the bifurcation scenario and cannot be linked to any `external'
orbit existing far away from the bifurcation.  This causes a generic
problem in the construction of the global uniform approximations which
will be discussed in section \ref{sec:uniform}.

In section \ref{sec:bridge}, we present some examples of bridge orbit
bifurcations in integrable and non-integrable systems with two degrees
of freedom.  In section \ref{sec:normalform}, we propose a new type of
normal form (derived in detail in the appendix) which successfully
describes the scenario of the bridge orbit bifurcation in integrable
systems.  In section \ref{sec:uniform}, we derive local and global
uniform approximations from this normal form, give analytic trace
formulae for the semiclassical density of states and compare its
numerically computed results with fully quantum-mechanical results for
the corresponding quantum Hamiltonians.  Section \ref{sec:summary} is
devoted to a summary and concluding remarks.

In the following sections, numerical integrations of the classical
equations of motion are performed with Adams' Method (a kind of
predictor-corrector method) to get solutions of high accuracy.
Periodic orbits are obtained by searching fixed points $(q,p)$ of the
Poincar\'e map in a suitable surface of section using a
two-dimensional Newton-Raphson iteration method, whereby the stability
matrix (linearized Poincar\'e map) is obtained at the same time.  The
fixed points are smooth functions of the system parameter and can
easily be followed through the bifurcation points under variation of
the parameter, as well as new branches of fixed points emerging from
the bifurcation points.

\section{Bridge orbit bifurcations in two dimensional $r^\alpha$
  potential models} \label{sec:bridge}

Let us start from a system with two degrees of freedom
described by the Hamiltonian
\begin{equation}
H_0(\bm{p},\bm{r})=\frac12\,\bm{p}^2+\frac12\,|\bm{r}|^\alpha.
\label{eq:hamil0}
\end{equation}
Since the potential is a homogeneous function of the coordinates
$\bm{r}=(x,y)$, the Hamiltonian has the scaling property
\begin{equation}
H_0(c^{\frac12}\bm{p},c^{\frac{1}{\alpha}}\bm{r})
  =cH_0(\bm{p},\bm{r})\,, \qquad c > 0\,,
\label{eq:scaleH}
\end{equation}
and the equations of motion (EOM) are invariant under the following
scaling transformation:
\begin{equation}
\bm{p}\to c^{\frac12}\bm{p},\quad
\bm{r}\to c^{\frac{1}{\alpha}}\bm{r},\quad
t\to c^{\frac{1}{\alpha}-\frac12}t\,,
\label{eq:scaling}
\end{equation}
while the energy transforms as $E\to cE$.  Therefore, the phase-space
profile is independent of energy and one has the same set of periodic
orbits at all energies $E$.  The action integral along a periodic
orbit (po) has the following energy dependence
\begin{eqnarray}
S_{\rm po}(E)=\oint_{{\rm po}(E)}\! \bm{p}\cdot \rmd\bm{r}
=\left(\frac{E}{E_0}\right)^{\!\!\frac12+\frac{1}{\alpha}}\!\!
\oint_{{\rm po}(E_0)}\!\bm{p}\cdot \rmd\bm{r}
=\hbar\,\cE\tau_{\rm po}\,,
\end{eqnarray}
where the dimensionless scaled energy $\cE$ and the dimensionless
scaled period $\tau_{\rm po}$ of the orbit are defined by
\begin{equation}
\cE=\left(\frac{E}{E_0}\right)^{\!\!\frac12+\frac{1}{\alpha}}\!\!,
    \qquad \tau_{\rm po}=\frac{1}{\hbar}\pp{S_{\rm po}}{\cE}
    =\frac{1}{\hbar}\oint_{{\rm po}(E_0)}\!\bm{p}\cdot \rmd\bm{r}\,,
\end{equation}
whereby the reference energy $E_0$ can be chosen arbitrarily.

\begin{figure}[p]
\begin{center}
\includegraphics[width=.6\linewidth]{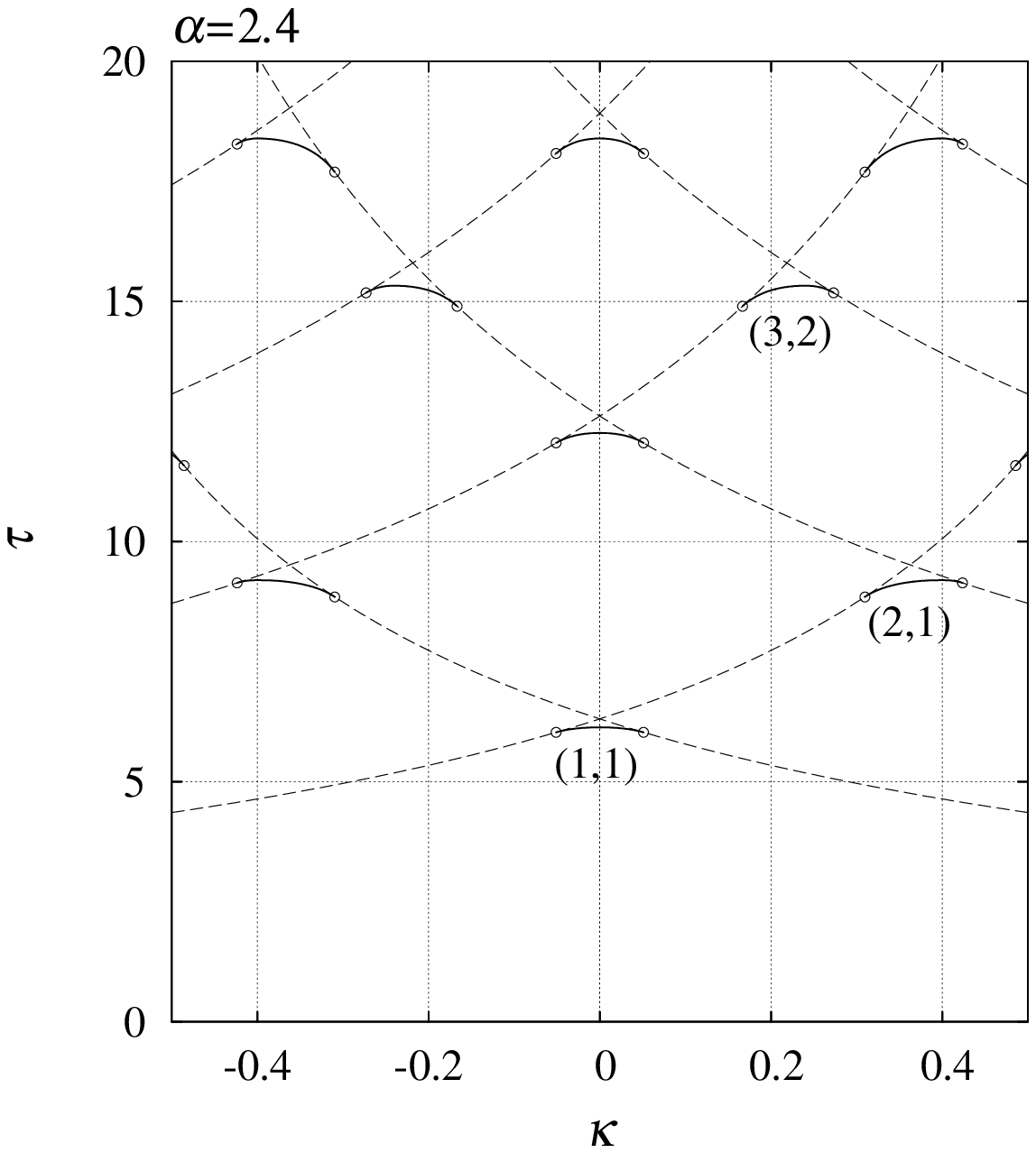}
\end{center}
\caption{\label{fig:bridge1}
Scaled periods $\tau$ of periodic orbits for the Hamiltonian
\eq{eq:hamil1} as functions of $\kappa$.  Isolated circular orbits
and families of bridge orbits are indicated by dashed and solid
curves, respectively.  Circles represent bifurcation points.}
\bigskip

\begin{center}
\includegraphics[width=\linewidth]{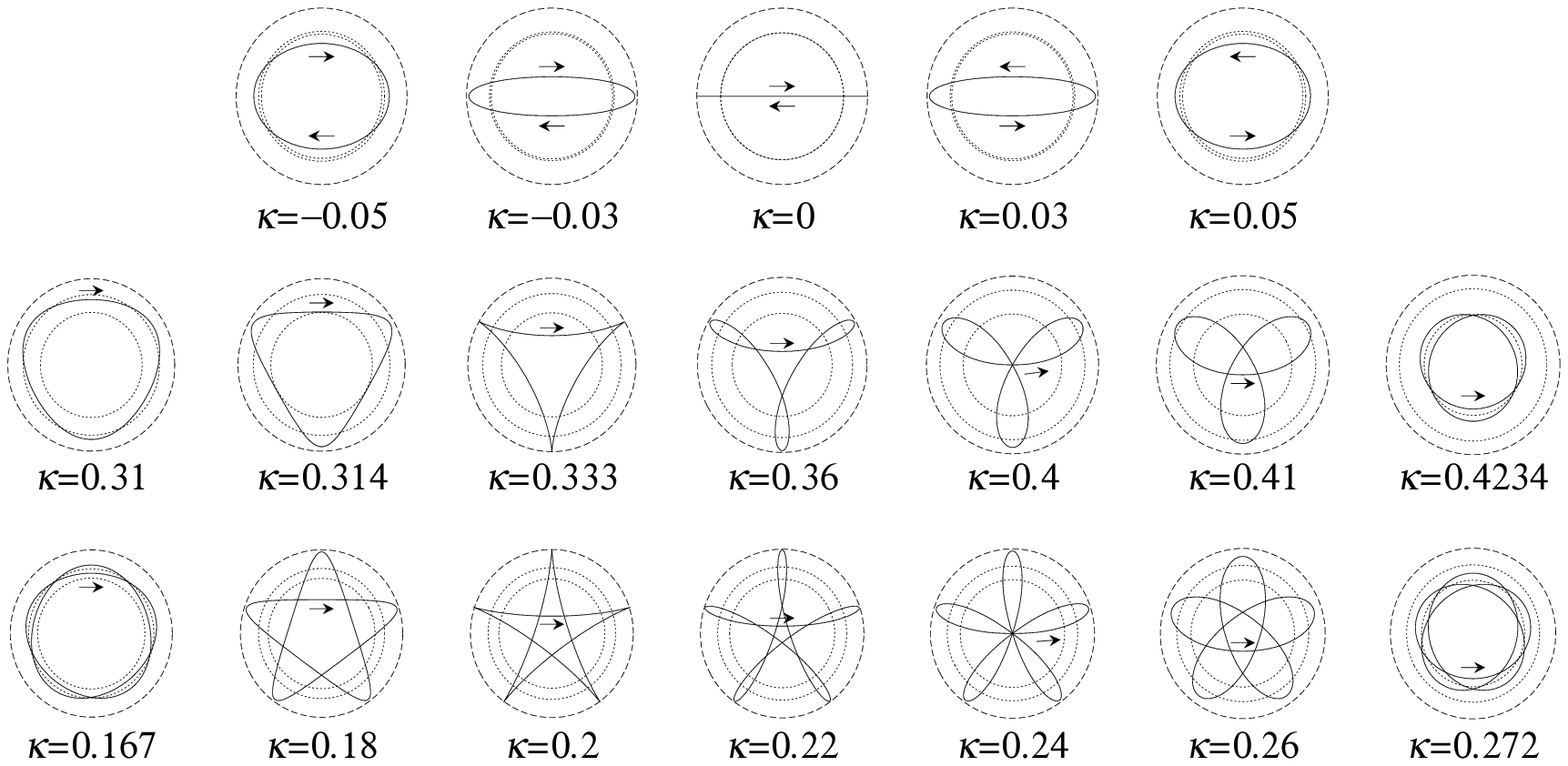}
\end{center}
\caption{\label{fig:po1}
Some short bridge orbits in Hamiltonian \eq{eq:hamil1} for
$\alpha=2.4$ and for different values of $\kappa$.  Upper row shows
symmetric (1,1) bridge, middle and lower rows show asymmetric (2,1)
and (3,2) bridges, respectively.}
\end{figure}

\begin{figure}[t]
\begin{center}
\includegraphics[width=\linewidth]{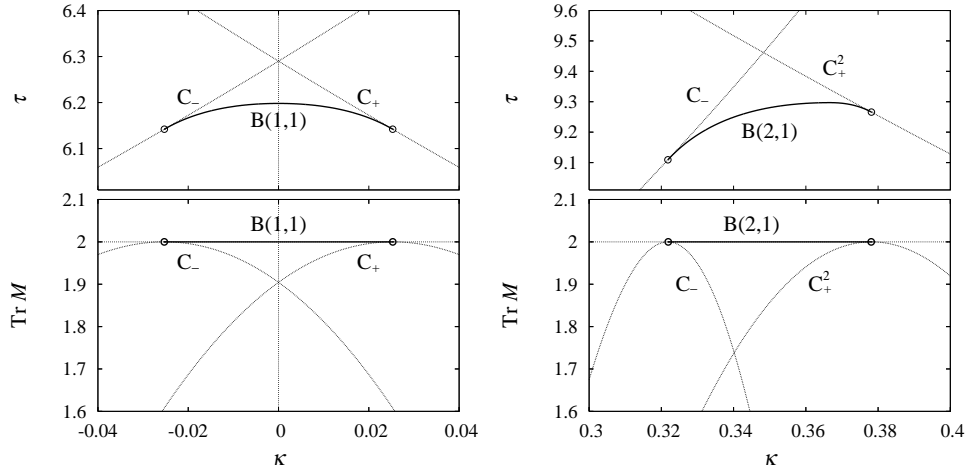}
\end{center}
\caption{\label{fig:varsk}
Scaled period $\tau$ (upper panels) and trace of stability matrix
$\mM$ (lower panels) of the symmetric (1,1) (left-hand side) and
asymmetric (2,1) (right-hand side) bridge orbits for the Hamiltonian
\eq{eq:hamil1} with $\alpha=2.2$, shown by solid lines as functions of
the parameter $\kappa$.  The dashed lines show the corresponding
quantities for the circular orbits C$_\pm$.}
\end{figure}

We now consider two kinds of perturbations of the system
\eq{eq:hamil0}.  The first one is introduced by a magnetic field
perpendicular to the $(x,y)$ plane.  The motion of a charged particle
in the plane is described by the Hamiltonian
\begin{equation}
H_\kappa=\frac12\,\bm{p}^2+\frac12\, r^\alpha
  -\kappa\, r^{\frac{\alpha}{2}-1}(xp_y-yp_x),
\label{eq:hamil1}
\end{equation}
where the radial dependence of the perturbation (with $r=|\bm{r}|$) is
determined such that the scaling invariance persists for any finite
$\kappa$.  The strength of the magnetic field is proportional to
$\kappa\,r^{\frac{\alpha}{2}-1}$.  Using polar coordinates
$\bm{r}=(r,\theta)$, $\bm{p}=(p_r,p_\theta)$, \eq{eq:hamil1} reads
\begin{equation}
H_\kappa=\frac12\left(p_r^2+\frac{p_\theta^2}{r^2}\right)
+\frac12\, r^\alpha-\kappa\, r^{\frac{\alpha}{2}-1}p_\theta\,.
\end{equation}
This system is integrable since $p_\theta=\Lambda$ is a constant of
motion.  For any non-zero $\kappa$, there exist two isolated circular
periodic orbits C$_\pm$ with different radii $r_\pm$ and angular
momenta $\Lambda_\pm$, which are found as solutions of the equations
\begin{eqnarray}
V_{\rm eff}(r,\Lambda)=\frac12\,r^\alpha+\frac{\Lambda^2}{2r^2}
-\kappa\Lambda r^{\frac{\alpha}{2}-1}=E\,, \label{effpot}\\
\pp{V_{\rm eff}}{r}(r,\Lambda)
=\frac{\alpha}{2}\,r^{\alpha-1}\!-\frac{\Lambda^2}{r^3}
-\kappa\Lambda\left(\frac{\alpha}{2}-1\right)r^{\frac{\alpha}{2}-2}=0\,.
\end{eqnarray}
Evidently they depend on the parameter $\kappa$.  The radii $r_\pm$
correspond simply to the minima of the effective potential
\eq{effpot}.  Small oscillations around the circular orbits C$_\pm$ 
have frequencies $\omega_\pm$ for their angular and $\Omega_\pm$
for their radial motions which are given by
\begin{eqnarray}
\omega_\pm=\frac{\Lambda_\pm}{r_\pm^2}-\kappa
r_\pm^{\frac{\alpha}{2}-1}, \qquad
\Omega_\pm=\sqrt{\pp{^2V_{\rm eff}}{r^2}(r_\pm,\Lambda_\pm)}\,.
\end{eqnarray}
They become periodic for those values $\kappa_\pm$ for which the 
two frequencies are commensurate, i.e., when
\begin{equation}
\frac{\Omega_\pm}{\omega_\pm}(\kappa_\pm)
=\pm\,\frac{(n_+ + n_-)}{n_\pm}\,,
\end{equation}
with positive integers $n_+$ and $n_-$.  Precisely at the values 
$\kappa=\kappa_\pm$, the orbits C$_\pm$ must undergo bifurcations, 
because the trace of their stability matrix $\mM$ equals two: 
$\TrM_{{\rm C}_\pm}(\kappa_\pm)=+2$.  The interesting phenomenon now
is that the two bifurcations are connected by a {\it bridge orbit} B
that is created (or absorbed) at the bifurcations.  It is actually a
degenerate family of orbits with $\TrM_{\rm B}(\kappa)=+2$ for all
values $\kappa_-\leq \kappa \leq \kappa_+$.  More precisely, a bridge
orbit B emerges from a bifurcation of the $n_-$-th repetition of the
orbit C$_-$ at $\kappa=\kappa_-$ and is absorbed at a bifurcation of
the $n_+$-th repetition of the orbit C$_+$ at $\kappa=\kappa_+$.  It
can therefore be labeled by the repetition numbers as B$(n_+,n_-)$.
Figure~\ref{fig:bridge1} shows the scaled periods of the shortest
periodic orbits of the system \eq{eq:hamil1} as functions of the
parameter $\kappa$.  At each crossing point of some repetitions of the
two circular orbits, indicated by the pair of numbers $(n_+,n_-)$, one
finds a bridge orbit family connecting them.  The shapes of some of
the bridge orbits are shown in figure~\ref{fig:po1}.  In
figure~\ref{fig:varsk}, the scaled periods and the traces of
stability matrices are plotted
as functions of $\kappa$ for the circular orbits C$_\pm$ and the
bridge orbits connecting them.  $\TrM_{{\rm C}_\pm}(\kappa)$
touches the horizontal line $\TrM=2$ at $\kappa=\kappa_\pm$, and in
between there exist the bridge orbit families having $\TrM_{\rm B}=2$.

\begin{figure}[p]
\begin{center}
\includegraphics[width=.6\linewidth]{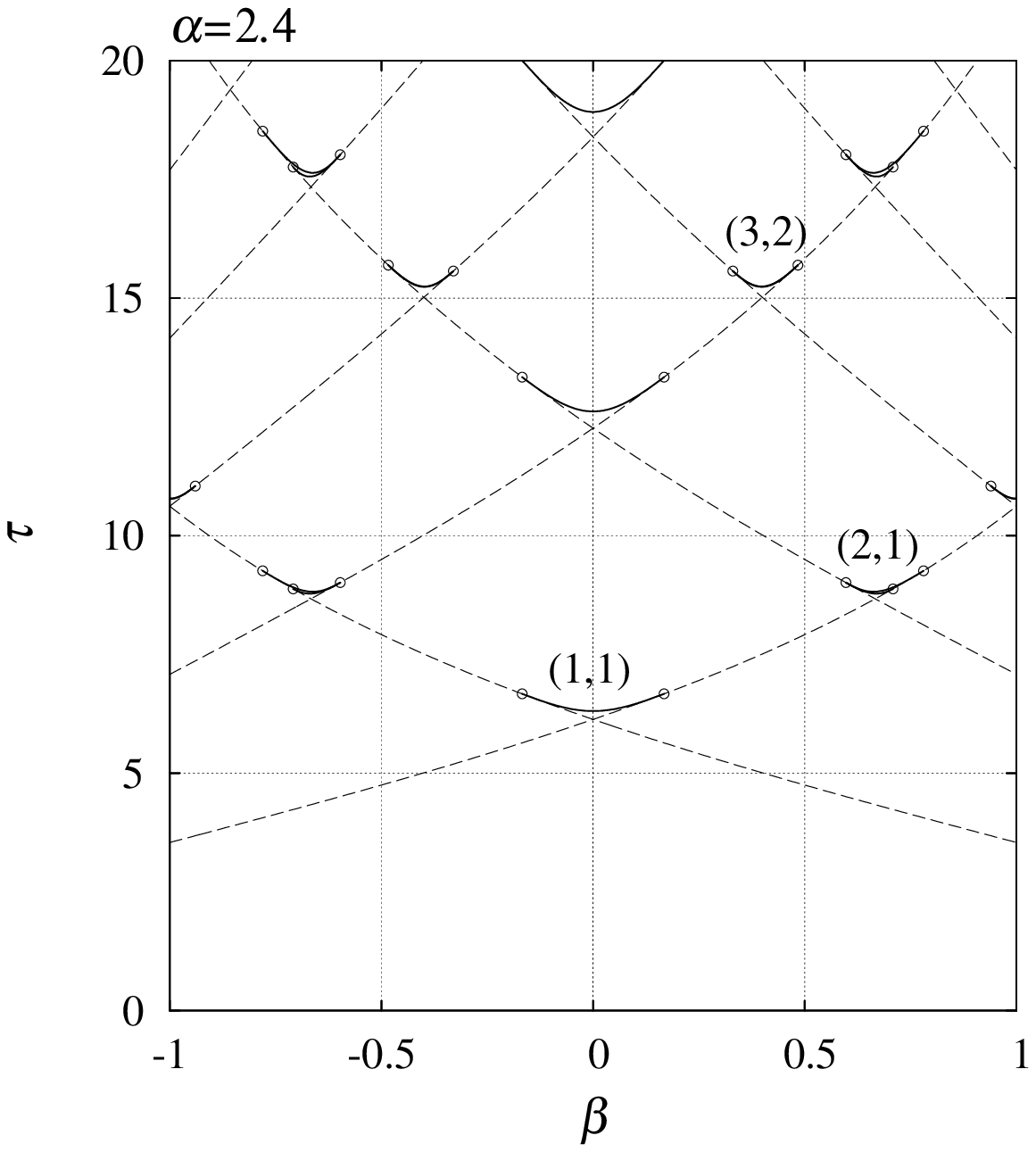}
\end{center}
\caption{\label{fig:bridge2}
Same as figure~\ref{fig:bridge1} but for the Hamiltonian
\eq{eq:hamil2} as functions of deformation parameter $\beta$.
Bridge orbits are isolated in contrast to those for
\eq{eq:hamil1}.}
\bigskip

\begin{center}
\includegraphics[width=.7\linewidth]{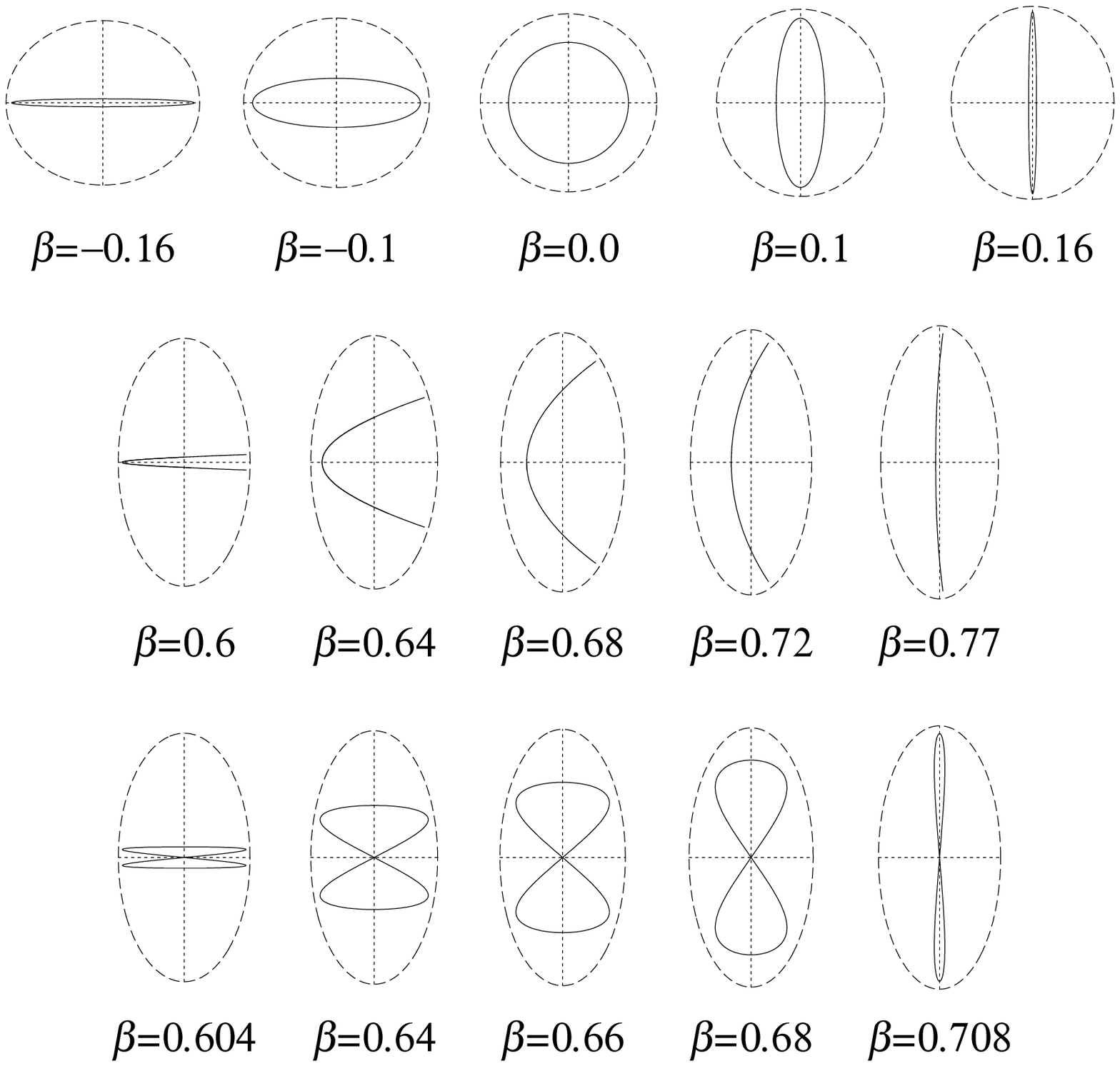}
\end{center}
\caption{\label{fig:po2}
Some short bridge orbits in the Hamiltonian \eq{eq:hamil1} for
$\alpha=2.4$ and for several values of deformation parameters $\beta$.
The upper row shows the symmetric bridge B(1,1), the middle and lower 
rows display the stable and unstable branches of the asymmetric 
B(2,1) bridge.}
\end{figure}

The second perturbation is introduced by an elliptic deformation.  
We modify the Hamiltonian \eq{eq:hamil0} as follows:
\begin{eqnarray}
H_\beta=\frac12\,\bm{p}^2+\frac12\left(rf_\beta(\theta)\right)^\alpha\!,
\label{eq:hamil2} \\
f_\beta(\theta)=\sqrt{\eta\cos^2\theta+\frac{1}{\eta}\sin^2\theta}\,,
\qquad \beta=\frac{2(\eta-1)}{\eta+1}\,.
\label{eq:beta}
\end{eqnarray}
The scaling rule \eq{eq:scaleH} persists for any $\beta$, but the
system is nonintegrable for $\beta\ne 0$.  We take $\beta$ as the
deformation parameter, which is related to the axis ratio $\eta$ by
the second equation in \eq{eq:beta}.  For any $\beta\ne 0$, there are
two isolated periodic orbits A$_x$ and A$_y$:
straight-line librations along the $x$ and $y$ axis, respectively.
Figure~\ref{fig:bridge2} shows the scaled periods of the shortest
periodic orbits in the system \eq{eq:hamil2} as functions of
$\beta$.  Again one, finds bridge orbits B which bifurcate from the
repetitions ($n_x,n_y$) of the orbits A$_x$ and A$_y$ near each
crossing point.

\begin{figure}
\begin{center}
\includegraphics[width=.5\linewidth]{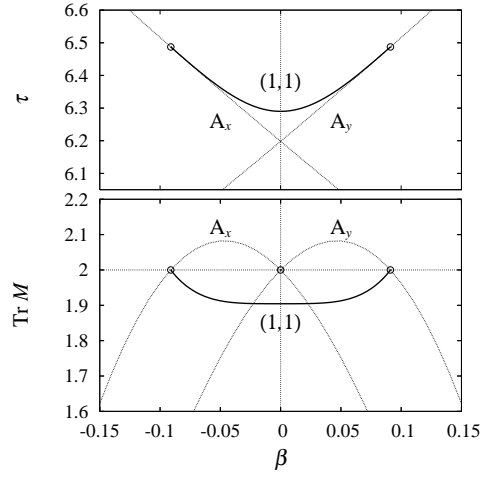}
\end{center}
\caption{\label{fig:varsb21}
Scaled period $\tau$ (upper panel) and trace of the stability
matrix $\mM$ (lower panel) of the symmetric (1,1) bridge
orbit for the Hamiltonian \eq{eq:hamil2} with $\alpha=2.2$,
shown by solid lines as functions of the deformation 
parameter $\beta$.  The dashed lines show the corresponding 
quantities for the A$_x$ and A$_y$ orbits.}
\end{figure}

\begin{figure}
\begin{center}
\includegraphics[width=\linewidth]{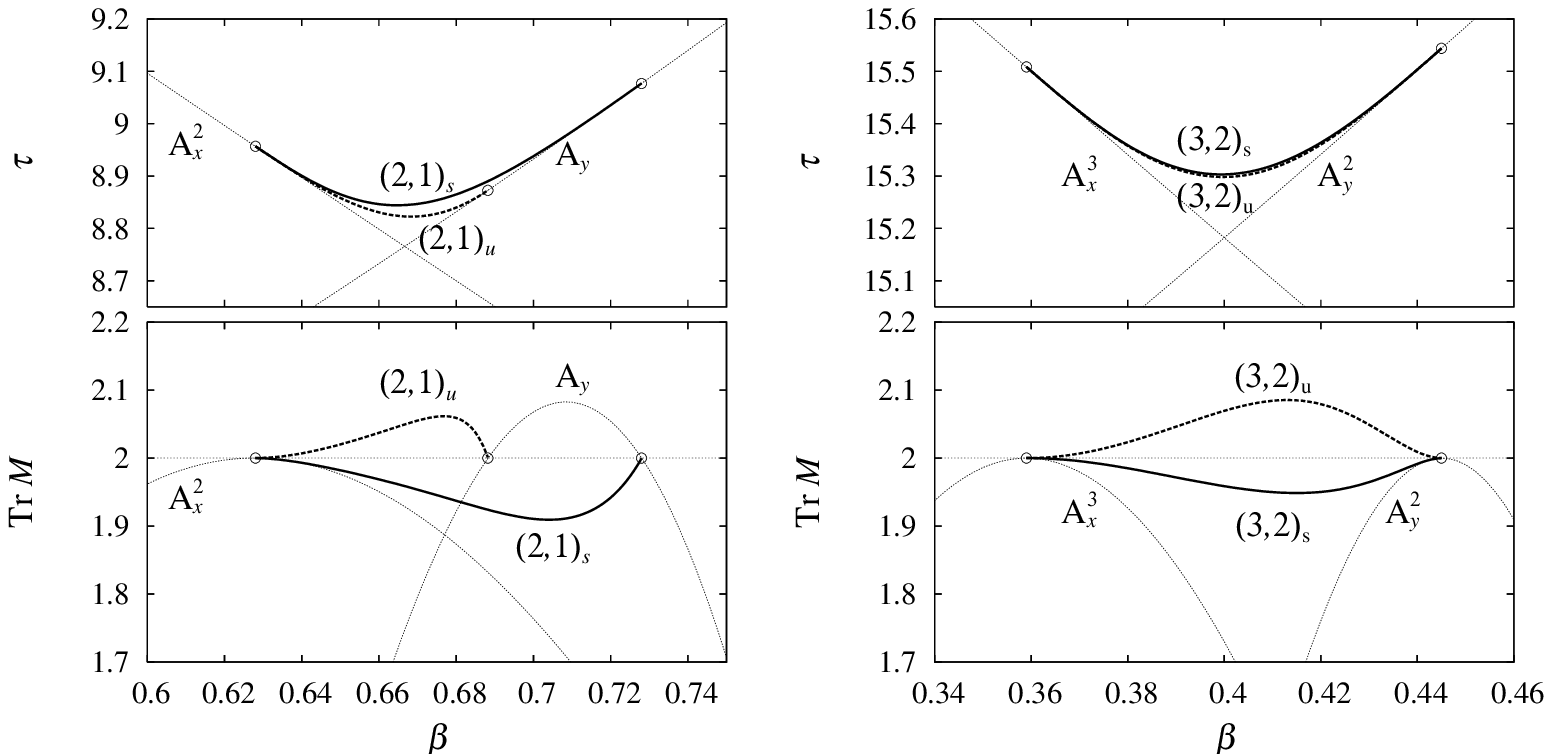}
\end{center}
\caption{\label{fig:varsb32}
Same as figure~\ref{fig:varsb21} but for asymmetric $(n_x,n_y)$ bridge
orbit bifurcations, with two pairs of values for the $n_x$-th
repetition of the A$_x$ orbit the and the $n_y$-th repetition of the
A$_y$ orbit.  The suffixes $s$ and $u$ represent the stable and
unstable branches, respectively.}
\end{figure}

In figure~\ref{fig:po2}, we plot some of the shortest
bridge orbits in the $(x,y)$ plane.  The system \eq{eq:hamil2} is
non-integrable and the bridge orbits here are isolated.  The symmetric
bridges B$(m,m)$ are bounded by two isochronous pitchfork
bifurcations; the example of B(1,1) is illustrated in
figure~\ref{fig:varsb21}.  In the left panel of
figure~\ref{fig:varsb32}, the graph of $\TrM(\beta)$ for the 2nd
repetition of the A$_x$ orbit touches the horizontal line $\TrM=+2$ at
$\kappa=0.628$ and two stable and unstable branches emerge from an
island-chain bifurcation.  The unstable and stable branches submerge
sequentially into the A$_y$ orbit at $\kappa=0.688$ and $0.728$,
respectively, in two successive pitchfork bifurcations.   For
asymmetric crossings $(n_x,n_y)$ with $n_x\neq n_y$, the bridges
consist of two isolated branches, i.e., one stable and one unstable
orbit, which we label by B$(n_x,n_y)_s$ and B$(n_x,n_y)_u$,
respectively.   For the bridge orbits B(2,1), the common left end is a
non-generic island-chain bifurcation of the 2nd repetition of the
A$_x$ orbit, while the right ends are two isochronous pitchfork
bifurcations of the A$_y$ orbit occurring at two different
deformations (see the left panel of figure~\ref{fig:varsb32}).  The
asymmetric B$(n_+,n_-)$ bridges with $n_+,n_-\geq 2$ also consist of
two isolated branches with different actions and stabilities, but
their end points are common at both ends of both branches, since they
occur both at period-multiplying island-chain bifurcations (cf. the
right panel in figure~\ref{fig:varsb32} for the B(3,2) bridges).

In the limit $\alpha\to 2$, $H_\kappa$ becomes the cranked and 
$H_\beta$ is the anisotropic harmonic oscillator.  (Note that both systems
have identical spectra after a suitable transfor\-mation from $\alpha$
to $\kappa$, cf. section 3.2.8 of \cite{BrackText}).  In this limit,
each bifurcation pair coalesces and the connecting bridge orbit 
shrinks to a point at which the two isolated orbits intersect.  At 
the crossing points, one has locally periodic-orbit families of 
two-fold degeneracy due to the dynamical SU(2) symmetry of the 
rationally deformed harmonic oscillator in two dimensions.

\begin{figure}[tb]
\begin{center}
\includegraphics[width=\textwidth]{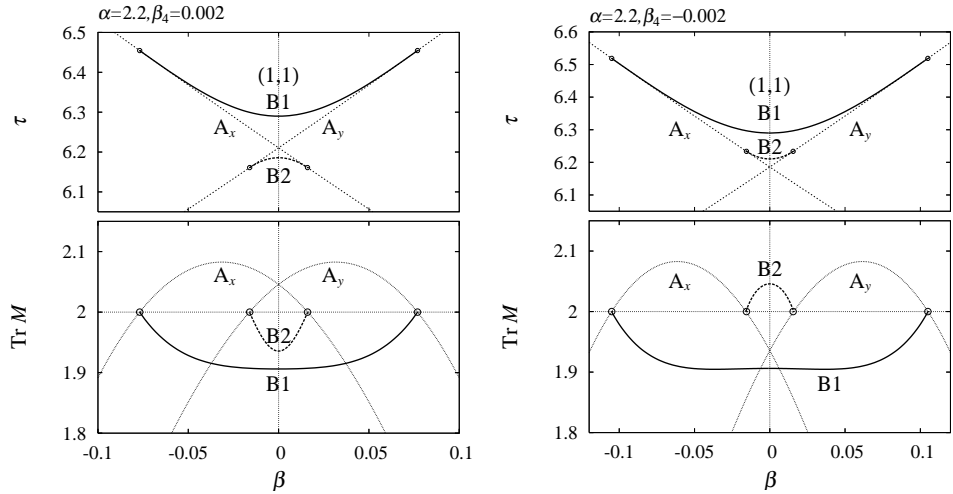}
\end{center}
\caption{\label{fig:varsb11w}
Same as figure~\ref{fig:varsb21} but for the shape function \eq{eq:beta4}
with $\beta_4=\pm 0.002$ and $\alpha=2.2$.}
\end{figure}

A scenario involving two bridge orbits is obtained if one breaks
the U(1) sym\-metry that the Hamiltonian \eq{eq:beta} possesses at 
$\beta=0$.  Let us e.g., modify the shape function $f_\beta(\theta)$ 
in \eq{eq:beta} by the following one:
\begin{equation}
f_{\beta,\beta_4}(\theta)=\sqrt{\eta\cos^2\theta
 +\frac{1}{\eta}\sin^2\theta}-\beta_4\cos4\theta\,,
\label{eq:beta4}
\end{equation}
with nonzero $\beta_4$.  In this case, a second bridge orbit appears
around $\beta=0$ for the symmetric $(m,m)$ bifurcations.
Figure~\ref{fig:varsb11w} shows the properties of these bridge orbits.
For $\beta_4=0$ it corresponds to figure~\ref{fig:varsb21} where the
two isolated orbits A$_x$ and A$_y$ intersect in a point with U(1)
symmetry $(\beta=0)$ and their stability traces intersect at $\TrM=2$
(see the lower part of the figure).  For $\beta_4\neq0$, this crossing
point is split, so that the second bifurcation of each orbit occurs at
different points on the $\beta$ axis; the new pair of bifurcations is
now connected by a second bridge orbit B2.  For $\beta_4>0$, as shown
on the left side of figure~\ref{fig:varsb11w}, both bridges are
stable, while for $\beta_4<0$ (see right side) one of them is stable
and the other is unstable.  All bifurcations here are of the
non-generic pitchfork type.

\begin{figure}[tb]
\begin{center}
\includegraphics[width=.75\textwidth]{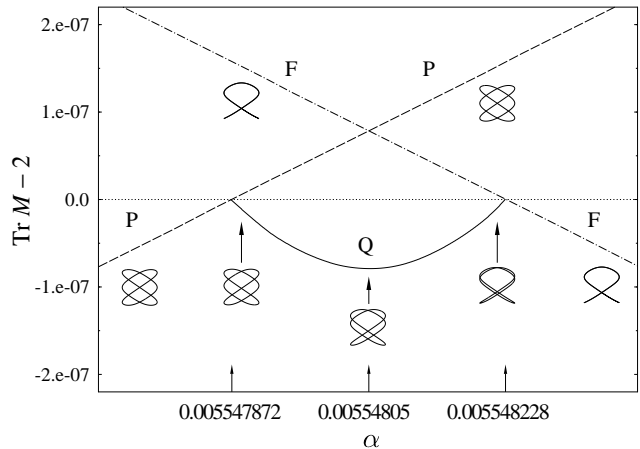}\hfill
\end{center}
\caption{\label{fig:erdabridge}
Stability exchange of isolated orbits F and P in the coupled
quartic oscillator \eq{q4} via two non-generic pitchfork 
bifurcations connected by an isolated bridge orbit Q.  The 
inserts exhibit the shapes of the orbits in the $(x,y)$ plane.}
\end{figure}

\medskip

Finally, we mention a bifurcation scenario that has been
discussed in \cite{ErikDahl}.  Hereby a pair of isolated orbits
exchange their stability via an isolated bridge orbit.  Examples 
for this are found in the coupled quartic oscillator
\begin{equation}
H_\alpha= \frac12\,\bm{p}^2 + \frac14\,(x^4+y^4)+\alpha\,x^2y^2\,.
\label{q4}
\end{equation}
Figure~\ref{fig:erdabridge} shows a narrow region of the 
chaoticity parameter $\alpha$.  The shapes of two crossing
isolated orbits F and P in the $(x,y)$ plane are shown by inserts, 
as well as various shapes of the bridge orbit Q interpolating 
between those two shapes.  Note the extremely small scale: the
maximum value of $|\TrM-2|$ of the bridge orbit is smaller than
10$^{-7}$.  On a larger scale, the bridge orbit may not be observed
numerically and the two isolated orbits F and P would appear to 
cross in a point.  The bifurcation diagram then would look similar 
to that of a transcritical bifurcation \cite{Transbif}.  The orbits
F and P are created at $\alpha=0.6315$ in a period-tripling 
bifurcation from a straight-line libration along the $y$ axis;
at $\alpha=0$ they become members of an integrable 3:2 torus with U(1)
symmetry.  We refer to \cite{Transbif} for details of this bifurcation
scenario and to \cite{ErikDahl,BraFeMaMi} for details of the potential
\eq{q4}.  A bridge bifurcation of the same type has been also found to
occur in a two-dimensional spin-boson Hamiltonian \cite{Plet}.

\section{Normal forms for some bridge-orbit bifurcation scenarios}
\label{sec:normalform}

Normal forms are frequently used in singularity theory (see
e.g., \cite{gols}) and catastrophe theory (see e.g., \cite{cata}) to
classify bifurcations.  The natural variables of the normal forms for
bifurcations in Hamiltonian systems with two degrees of freedom are
the two canonical variables $(q,p)$ spanning a projected Poincar\'e
surface of section transverse to the bifurcating parent orbit, and a
bifurcation parameter $\epsilon$.  These variables are usually chosen
such that the bifurcation of the parent orbit occurs at
$(q,p,\epsilon)= (0,0,0)$.  The normal form function $S(q,p,\epsilon)$
must fulfill the condition that its critical points correspond to the
fixed points in $(q,p,\epsilon)$ space in the neighborhood of the
origin, and hence to the periodic orbits taking part in the
bifurcation.  This condition, however, is not sufficient to specify
the normal form for a given bifurcation uniquely.  Usually,
$S(q,p,\epsilon)$ is chosen to contain a minimum number of parameters
and simple functions of the variables $(q,p,\epsilon)$ -- often just
polynomial expressions in $q$, $p$ and $\epsilon$ -- that yield the
desired fixed-point scenario of a given bifurcation.  For the generic
bifurcations in two-dimensional symplectic maps according to the
classification of Meyer \cite{Meyer}, the standard normal forms have
been given in \cite{Ozorio87,Ozobook}.  They can also be used for
non-generic bifurcations of the same type (i.e., with the same
fixed-point scenario in phase space) occurring in systems with
discrete symmetries (see e.g., \cite{nong}).  Normal forms for
non-generic bifurcations in Hamiltonian systems
with different fixed-point scenarios can be found in \cite{Transbif} (for
the transcritical bifurcation) and in \cite{Schomerus97b,Kaidel04a}
(for codimension-two bifurcations).

For bifurcations involving more than two orbits, it is often useful 
to transform the Poincar\'e variables $(q,p)$ to action-angle 
variables $(\varphi,I)$ (cf. \cite{Ozobook,Sieber98}):
\begin{equation}
p=\sqrt{2I}\cos\varphi\,, \quad q=\sqrt{2I}\sin\varphi\,,
\qquad I\geq0\,,\quad \varphi\in[0,2\pi)\,.
\label{standmap}
\end{equation}
This becomes particularly useful when one considers integrable systems
for which the normal form function does not depend on the angle
$\varphi$, so that one only has to deal with a function
$S(I,\epsilon)$ depending on two variables.  For instance, in
\cite{Kaidel04a}, a simple bifurcation of a torus from an isolated
orbit in the integrable Hamiltonian
$H=(p_x^2+p_y^2)/2+(x^2+y^2)/2-\lambda\,y^3\!/3$ could be described by
the normal form
\begin{equation}
S_1(I,\epsilon)=S_0-\epsilon I+aI^2
\label{eq:nform1}
\end{equation}
with suitably chosen constants $S_0$ and $a$ and bifurcation parameter
$\epsilon$.  The situation there corresponds to one half of that
seen in figure~\ref{fig:varsk}: a torus B bifurcates from an isolated
orbit C.  The stationary condition for \eq{eq:nform1} is
\begin{equation}
\pp{S_1}{I}(I_{\rm B},\epsilon)=0\,, \quad\Rightarrow\quad
I_{\rm B}=\frac{\epsilon}{2a}\,.
\label{statS1}
\end{equation}
Since $I$ must be positive definite, the torus B exists only for
$\epsilon/a>0$; its stability trace is always $\TrM_{\rm B}=+2$ there.
The isolated parent orbit arises from $I=0$ as a semiclassical 
end-point correction (see \cite{Kaidel04a} for details).  Its stability 
trace is given by $\TrM_{\rm C}=2-\epsilon^2$, so that it is always stable
except at the bifurcation point $\epsilon=0$.

In this section, we develop normal forms for some of the 
bridge-bifurcation scenarios described in the previous section.
We start with the integrable model \eq{eq:hamil1}.
\begin{figure}
\begin{center}
\includegraphics[width=.4\textwidth]{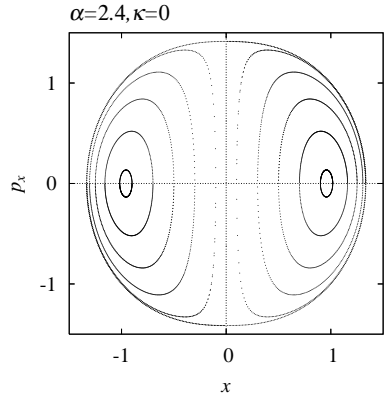}
\end{center}
\caption{\label{fig:pss}
Poincar\'e surface of section for \eq{eq:hamil1} with
$\alpha=2.4$ and $\kappa=0$.}
\end{figure}
Figure~\ref{fig:pss} shows a Poincar\'e surface of section obtained 
for the parameters $\alpha=2.4$ and $\kappa=0$, fixing $y=0$ (and 
$\dot{y}>0$).  Since the system is integrable, all fixed points lie on 
continuous curves which are intersections of rational tori with 
the $(x,p_x)=(q,p)$ plane, except for the two isolated fixed points 
along the $x$ axis which belong to the isolated rotating
orbits C$_\pm$.  The corresponding bifurcation diagrams are seen in 
figure~\ref{fig:bridge1} where the rational tori correspond to the 
B(1,1) bridge orbit.  The envelope of the curves in figure~\ref{fig:pss}
corresponds to the boundary of the classically accessible phase
space given by $p_y=0$.

The phase-space diagram seen in figure~\ref{fig:pss} is completely
analogous to that of a two-dimensional isotropic harmonic oscillator
$H=(p_x^2+p_y^2)/2+\omega^2(x^2+y^2)/2$.  The tori correspond to the
elliptic orbits, the isolated points to the circular orbits C$_\pm$
(with both time orientations), and the line $x=0$ to the librating
orbits.  (The boundary contains the librating orbits along the $y$
axis, i.e., with $p_y=0$, which strictly cannot be seen in the
Poincar\'e plot.) It is useful to parameterize the orbits by their
(conserved) angular momentum $L=xp_y-yp_x$.  The circular orbits C$_+$
and C$_-$ have maximum and minimum value of $L$, respectively, and all
intermediate nonzero $L$ values correspond to elliptic orbits which
are degenerate against rotations around the origin about an angle
$\phi\in[0,\pi)$.  $L=0$ corresponds to the librating orbits which
have the same degeneracy.

This analogy suggests to use the angular momentum $L$ and the 
rotation angle $\phi$ of the harmonic-oscillator orbits to define 
action-angle variables $\varphi=2\phi$ and $I=L/2$ and map these 
onto the Poincar\'e variables of our present integrable system 
\eq{eq:hamil1}.  This transformation is derived in \ref{app:mapping}. 
The resulting mapping of $(\varphi,I)$ to the Cartesian Poincar\'e 
variables $(q,p)$ is given by the relations (here for $\omega=1$)
\begin{eqnarray}
q&=\sqrt{2\rho}\,\frac{\cos\theta}{\sqrt{1-\sin\theta\cos\varphi}}\,,& \NN
p&=\sqrt{2\rho}\,\frac{\sin\theta\sin\varphi}{\sqrt{1
   -\sin\theta\cos\varphi}}\,,
 &\rho>0\,,\quad 0\leq\varphi\leq 2\pi\,,\NN
I&=\frac12\,q\sqrt{4\rho-q^2-p^2}=\rho\cos\theta\,,\qquad
 & 0 \leq \theta \leq \pi\,,
\label{eq:mapping}
\end{eqnarray}
which define an area-conserving canonical transformation
$(\varphi,I)\leftrightarrow(q,p)$.  Hereby $\theta$ is an angle 
parameterizing the angular momentum by $L=L_c\cos\theta$, where 
$L_c$ is the angular momentum of the C$_+$ orbit in 
the harmonic oscillator.  Note that the transformation defined by
\eq{eq:mapping} is more complicated than that given in 
\eq{standmap}; in particular, the action variable $I$ here can have
both signs.  It is limited by the values $\pm\rho$ yielding the fixed 
points $(q,p)=(\pm\sqrt{2\rho},0)$ which correspond to the two
isolated orbits C$_\pm$ at their bifurcation points where the bridge
orbit is created or absorbed.

It turns out now that the bridge bifurcations in the
integrable system \eq{eq:hamil1} can be described by the same
normal form function $S_1(I,\epsilon)$ as defined in \eq{eq:nform1}
above, except that here we have to use the definition of $I$ in
\eq{eq:mapping}.
As shown in \ref{app:mapping}, $\rho$ equals $L_c/2$ in the 
harmonic oscillator model.  Here, $\rho>0$ is simply a parameter
of the normal form that will be determined in section \ref{sec:uniform}.

The stationary points of $S_1(I,\epsilon)$ are most conveniently
found in terms of the variable $\theta$.  The stationary condition is
\begin{equation}
\pp{S_1}{\theta}=\rho\sin\theta(\epsilon-2aI)=0\,.
\end{equation}
The stationary points satisfying $\sin\theta=0$, i.e., $\theta=0$ and
$\pi$, correspond to the two isolated circular orbits C$_\pm$ with 
$I=\pm\rho$.  In addition, there is another stationary point 
satisfying $I_{\rm B}=\epsilon/2a$ as in \eq{statS1} above, corresponding 
to the B torus.  Since $I_{\rm B}$ must also fulfill the condition 
$I_{\rm B}=\rho\cos\theta$, it has real values only for
\begin{equation}
-2|a|\rho\leq\epsilon\leq 2|a|\rho\,,
\end{equation}
so that the B torus only exists in the range of $\epsilon$ values
between the bifurcation points of the C$_\pm$ orbits.  The actions of 
these periodic orbits are obtained by inserting their stationary
values of $I$ into the normal form \eq{eq:nform1}; they become
\begin{eqnarray}
S_{{\rm C}_\pm}=S_1(\pm\rho,\epsilon)=S_0\mp\epsilon\rho+a\rho^2, \quad
S_{\rm B}=S_1(I_{\rm B},\epsilon)=S_0-\frac{\epsilon^2}{4a}.
\label{eq:action1}
\end{eqnarray}
The traces of their stability matrices can be obtained from the 
normal form by \cite{Sieber98}
\begin{eqnarray}
\TrM=\left(
 \frac{\partial^2\hat{S}}{\partial p\partial q}
 \right)^{\!\!-1}\!\left[1+\left(
 \frac{\partial^2\hat{S}}{\partial p\partial q}
 \right)^{\!\!2} \! -
 \frac{\partial^2\hat{S}}{\partial p^2}
 \frac{\partial^2\hat{S}}{\partial q^2}
 \right]\!,\NN
\hat{S}(q,p,\epsilon)=S_1(q,p,\epsilon)+qp\,,\label{genfunc}
\end{eqnarray}
whereby $\hat{S}(q,p,\epsilon)$ is the generating function of the 
Poincar\'e map with initial momentum $p$ and final coordinate $q$ 
in the $(q,p)$ plane at the value $\epsilon$ of the bifurcation
parameter.  From this we obtain for our periodic orbits
\begin{equation}
\TrM_{{\rm C}_\pm} = 2-(\epsilon\mp 2a\rho)^2\,, \qquad
\TrM_{\rm B}=2\,.
\label{eq:trm1}
\end{equation}
Hence we see that the normal form \eq{eq:nform1} with the mapping 
\eq{eq:mapping} correctly describes the bifurcation scenario
of the bridge orbits found in the model \eq{eq:hamil1}, as illustrated 
in figure~\ref{fig:bridge1}.

In non-integrable systems, the normal form must depend also on the
angle variable $\varphi$.  Let us consider the following normal form
\begin{eqnarray}
S_2(I,\varphi,\epsilon)&=&S_0-\epsilon I+aI^2+b(\rho^2-I^2)\cos^2\varphi \NN
&=&S_0-\epsilon\rho\cos\theta
   +\rho^2(a\cos^2\theta+b\sin^2\theta\cos^2\varphi)\,,
\label{eq:nform2}
\end{eqnarray}
which respects the fact that the $\varphi$-dependent terms in any 
canonically invariant quantity (such as action, stability) should 
vanish for $I=\pm\rho$.  The stationary phase conditions become
\begin{subequations}
\begin{eqnarray}
\pp{S_2}{\varphi}&=&-2b\rho^2\sin^2\theta\sin\varphi\cos\varphi=0\,,\\
\pp{S_2}{\theta}&=&\rho\sin\theta(\epsilon-2\rho\cos\theta
 (a-b\cos^2\varphi))=0\,.
\end{eqnarray}
\end{subequations}
The above set of equations have the following solutions:
\begin{subequations}
\begin{eqnarray}
\sin\theta=0\,, \quad \theta_{\pm}=0\,,~\pi \quad\Leftrightarrow\quad I_\pm=\pm\rho\,,
  \label{eq:ni_ends}\\
\cos\varphi=0\,,\quad \varphi_{\rm B1}=\frac{\pi}{2},~\frac{3\pi}{2} \quad\Leftrightarrow\quad
  I_{\rm B1}=\rho\cos\theta_{\rm B1}=\frac{\epsilon}{2a}\,,
  \label{eq:ni_brd1}\\
\sin\varphi=0\,, \quad
  \varphi_{\rm B2}=0\,,~\pi \quad\Leftrightarrow\quad
  I_{\rm B2}=\rho\cos\theta_{\rm B2}=\frac{\epsilon}{2(a-b)}\,.
  \label{eq:ni_brd2}
\end{eqnarray}
\end{subequations}
The periodic orbits corresponding to the solutions \eq{eq:ni_brd1} and
\eq{eq:ni_brd2} exist for $-2|a|\rho<\epsilon<2|a|\rho$ and
$-2|a-b|\rho<\epsilon<2|a-b|\rho$, respectively.  The actions
of these periodic orbits are
\begin{subequations}\label{eq:action2}
\begin{eqnarray}
S_\pm=S_0\mp\epsilon\rho+a\rho^2, \\
S_{\rm B1}=S_0-\frac{\epsilon^2}{4a}\,, \\
S_{\rm B2}=S_0-\frac{\epsilon^2}{4(a-b)}+b\rho^2,
\end{eqnarray}
\end{subequations}
and the traces of their stability matrices are
\begin{subequations}\label{eq:trm2}
\begin{eqnarray}
\TrM_\pm=2-(\epsilon\mp 2a\rho)[\epsilon\mp 2(a-b)\rho]\,, \\
\TrM_{\rm B1}=2+\frac{b}{a}\,(\epsilon-2a\rho)(\epsilon+2a\rho)\,,\\
\TrM_{\rm B2}=2-\frac{b}{a-b}\,[\epsilon-2(a-b)\rho][\epsilon+2(a-b)\rho].
\end{eqnarray}
\end{subequations}
Thus, the periodic orbits \eq{eq:ni_brd1} and \eq{eq:ni_brd2} are the two
bridge orbits which connect the two periodic orbits corresponding to
\eq{eq:ni_ends} at the bifurcation points $\epsilon_{\rm bif}=\pm 2a\rho$
and $\pm 2(a-b)\rho$.  The bridge orbits found in figure~\ref{fig:varsb11w} 
can be regarded as of this type.  For $b=a$ (or $a=0$) the bridge orbit
B2 (or B1) shrinks to a single point $\epsilon=0$.  This corresponds to
the situation found for the (1,1) orbit in figure~\ref{fig:varsb21}.
The orbit Q in figure~\ref{fig:erdabridge}
can be regarded as the bridge orbit B2 in the limit of extremely small
$|a-b|$.

\begin{figure}[p]
\begin{center}
\includegraphics[width=\linewidth]{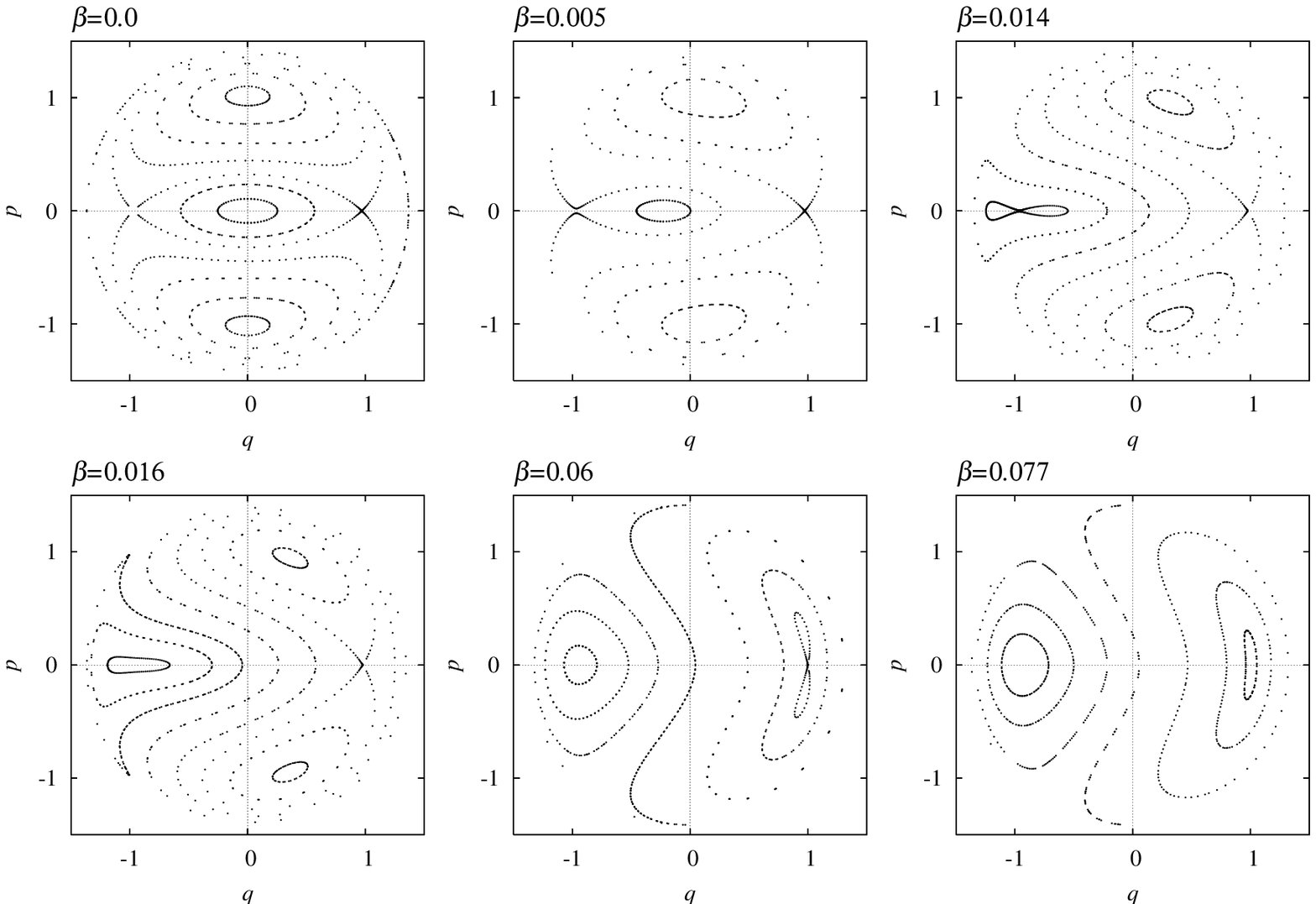}
\end{center}
\vspace{-\baselineskip}
\caption{\label{fig:pmap2}
Poincar\'e surface of section for the Hamiltonian \eq{eq:hamil2} with
deformation \eq{eq:beta4}, plotted for $\alpha=2.2$, $\beta_4=0.002$
and several values of $\beta$ in the (1,1) bifurcation region.
(See text for definitions of the Poincar\'e variables $(q,p)$.)}
\bigskip

\begin{center}
\includegraphics[width=\linewidth]{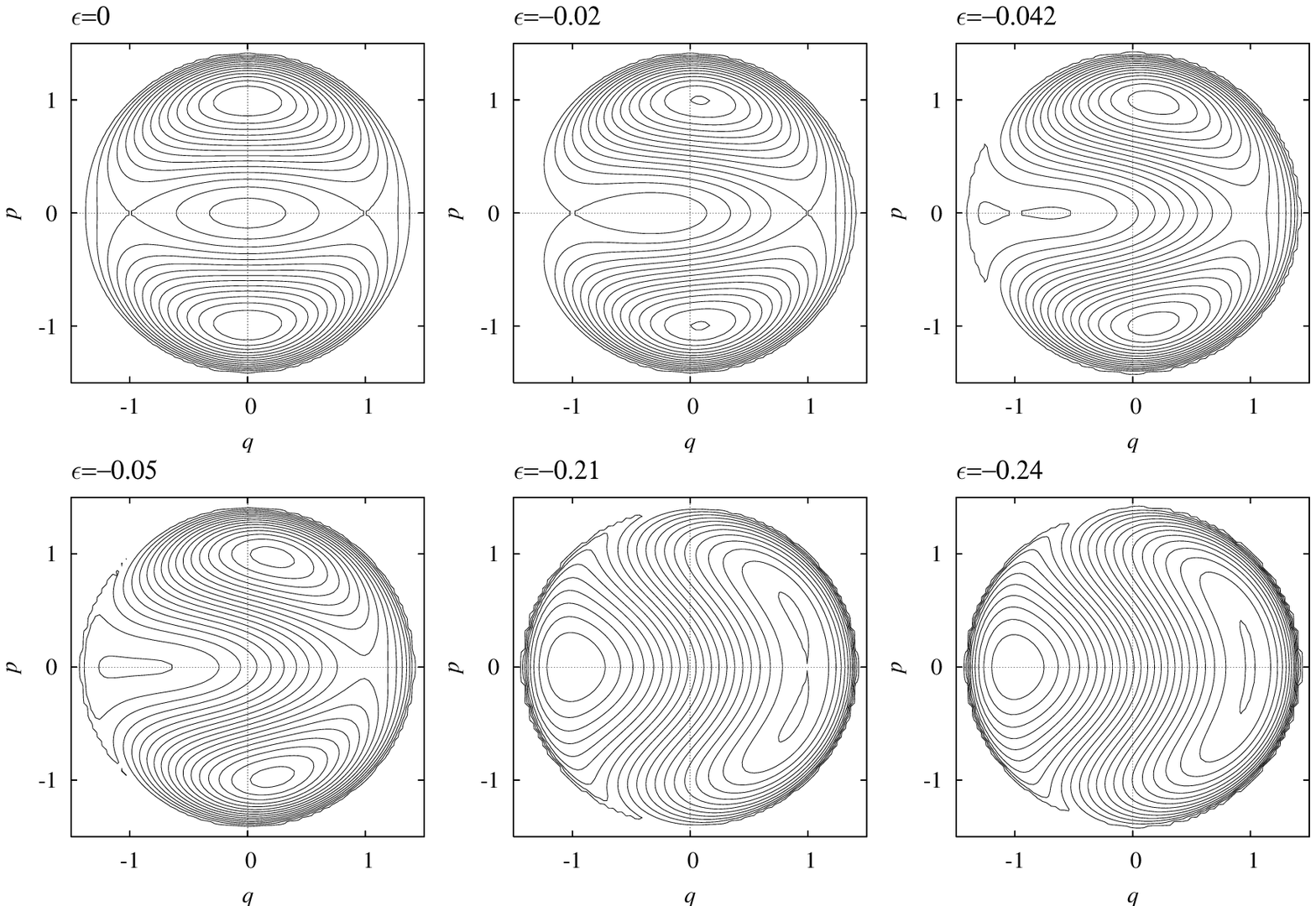}
\end{center}
\vspace{-\baselineskip}
\caption{\label{fig:cmap2}
Contour map of the normal form \eq{eq:nform2} with $a\rho=-0.12$,
$b\rho=-0.145$ and several values of $\epsilon$ in bifurcation region
corresponding to each panels of figure~\ref{fig:pmap2}.
$(q,p)$ are plotted in units of $\sqrt{2\rho}$.}
\end{figure}

Figure~\ref{fig:pmap2} shows the Poincar\'e surface of section
for the Hamiltonian \eq{eq:hamil2}, with deformations given by
\eq{eq:beta4}, for $\alpha=2.2$, $\beta_4=0.002$ and several values 
of $\beta$ along the right half of the (1,1) bridge bifurcation region
(cf. the left panel of figure~\ref{fig:varsb11w}).  The Poincar\'e
variables $(q,p)$ are here defined by $q=(x+y)/\sqrt{2}$,
$p=(p_x+p_y)/\sqrt{2}$ at the surface of section $x-y=0$,
$\dot{x}-\dot{y}>0$.  At $\beta=0$, the existing periodic orbits are
the bridge B1 at $(q,p)\approx(0,\pm1)$, the bridge B2 at
$(q,p)=(0,0)$, and the diameter orbits A$_x$, A$_y$ at
$(q,p)\approx(\mp1,0)$.  Both bridges are stable and the diameters are
unstable.  At the bifurcation point $\beta=0.016$, the bridge B2
merges into the A$_x$ orbit and A$_x$ becomes stable.  At the
bifurcation point $\beta=0.077$, the bridge B1 merges into the A$_y$
orbit and A$_y$ becomes stable.
By fitting \eq{eq:action2} and \eq{eq:trm2} to the periodic orbit
quantities for $\alpha=2.2$ and $\beta_4=0.002$, we obtain
$a\rho=-0.12$ and $b\rho=-0.145$.  The bifurcation points are thus
$\epsilon=\pm0.24$ and $\pm0.05$ for the B1 and B2 bridges,
respectively.  Figure~\ref{fig:cmap2} shows the contour map of the
normal form \eq{eq:nform2} with the above values of $a$ and $b$,
and with $\epsilon$
corresponding to each value of $\beta$ in figure~\ref{fig:pmap2}.  One
will see that the phase space profiles of figure~\ref{fig:pmap2} are
nicely reproduced by this normal form.

For the asymmetric bridges $(n_+,n_-)$ with $n_+\ne n_-$ we have so
far not found any suitable normal form.  One may need a mapping based
on the anisotropic (rational) harmonic oscillator instead of that
given in \eq{eq:mapping} which is based on the isotropic oscillator.
This will be investigated in further research.

\section{Uniform approximations for the density of states}
\label{sec:uniform}

\subsection{Local uniform approximation}

In this section we evaluate the semiclassical level density around the
bifurcation points using the normal form obtained in the above
section.  In the following we limit ourselves to the integrable model
described by the Hamiltonian \eq{eq:hamil1}.  The calculation of the
parameters in a normal form from a given Hamiltonian is in general a
very difficult problem.  The idea of the uniform approximations is to
avoid their direct calculation by relating them to the local invariant
properties of the participating periodic orbits in the vicinity of a
bifurcation, which can be obtained numerically from a periodic-orbit
search.  The bridge bifurcations occurring in the Hamiltonian
\eq{eq:hamil1} involve three orbits: the two circular orbits C$_\pm$
(with repetition numbers $n_+$ and $n_-$) and the bridge orbits
B$(n_+,n_-)$ (cf. figures~\ref{fig:bridge1}, \ref{fig:varsk}).  From
these, we can determine five independent quantities: the three actions
$S_\pm$ and $S_{\rm B}$, and the two stability traces $\TrM_\pm$
(recall that $\TrM_{\rm B}=2$ is constant).  The normal form given by 
\eq{eq:nform1} and \eq{eq:mapping} contains the four parameters $S_0$, 
$\epsilon$, $a$ and $\rho$, 
which we can determine using four 
of the above five orbit properties.  In problems with only one
bifurcation point, such a way of determining the normal form
parameters does not contradict with remaining unused quantities, 
but in the present situation of the bridge bifurcations, the values 
of the parameters do depend on which quantities are used.  This 
problem is related with the global nature of the bridge bifurcations, 
where the parameters undergo significant changes between the two
bifurcation points.  Different from all bifurcations treated so
far in the literature in uniform approximations, the bridge orbit
here has no `external link', i.e., there is no external orbit
outside the bifurcation interval to which its properties can be
asymptotically linked.  This leads to a slight ambiguity in determining
the parameter $a$, as we shall see below.

In order to determine the normal form parameters uniquely, at least 
locally for each given $\epsilon$, we add one more term 
to $S_1(I,\epsilon)$ in \eq{eq:nform1} and use the normal form
\begin{equation}
S_3(I,\epsilon)=S_0+\epsilon I+aI^2+bI^3, \qquad
I=\rho\sin\theta\,. \label{eq:nform3}
\end{equation}
The stationary-phase analysis then predicts the properties of the
periodic orbits to be
\begin{subequations}\label{eq:nfparam}
\begin{eqnarray}
S_\pm=S_0+a\rho^2\mp(\epsilon\rho-b\rho^3) \\
S_{\rm B}=S_0-\frac{\epsilon^2(1+2b\epsilon/a^2+\sqrt{1+3b\epsilon/a^2})}{
  a(1+\sqrt{1+3b\epsilon/a^2})^3} \\
\TrM_\pm=2-[\epsilon-(\pm2a\rho+3b\rho^2)]^2.
\end{eqnarray}
\end{subequations}
The five normal form parameters ($S_0$, $\epsilon$, $a$, $b$, and
$\rho$) are now uniquely determined by the five equations in 
\eq{eq:nfparam}, although these cannot be solved analytically.
Due to the scaling rules, the parameters have the following
energy dependences:
\begin{equation}
S_0=\hbar\cE\tau_0\,, \quad
\rho=\hbar\cE\tilde{\rho}\,, \quad
a=\frac{\tilde{a}}{\hbar\cE}, \quad
b=\frac{\tilde{b}}{(\hbar\cE)^2}\,,
\label{eq:dlp}
\end{equation}
where $\tau_0$, $\tilde{\rho}$, $\tilde{a}$, and $\tilde{b}$ are
dimensionless constants. 
Ideally, these four parameters should not depend on the bifurcation 
parameter $\epsilon$ throughout the bifurcation region.

\begin{figure}[t]
\begin{center}
\includegraphics[width=\textwidth]{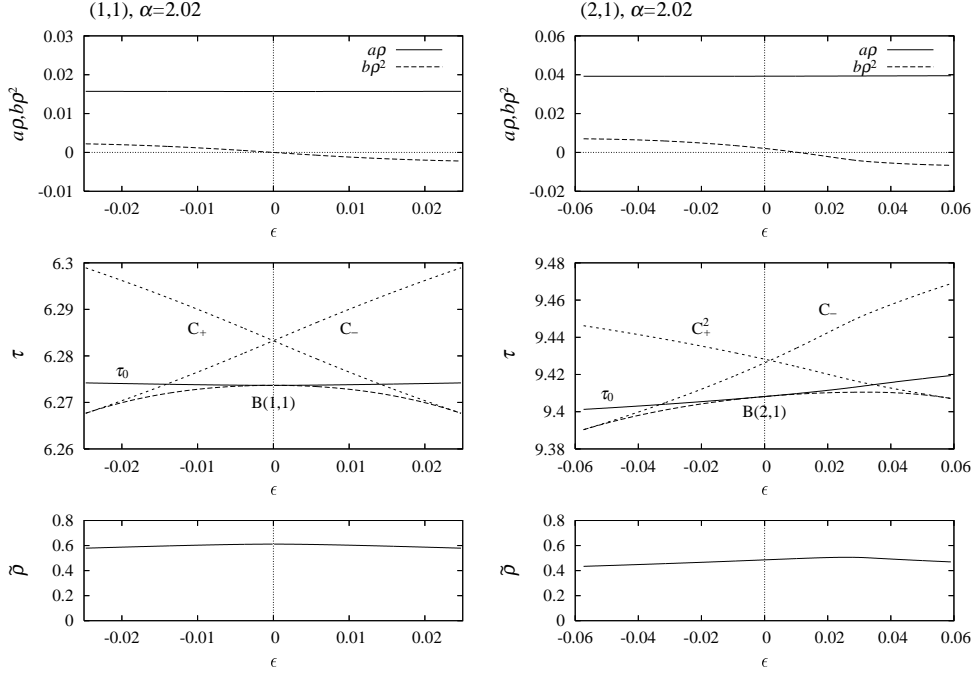}
\caption{\label{fig:nfparam}
Normal form parameters determined from periodic orbit
quantities.  Left and right panels are results for the
(1,1) and (2,1) bridge bifurcations, respectively,
for $\alpha=2.02$.  In the center panels, the scaled periods 
$\tau$ of all three orbits are shown besides 
$\tau_0=S_0/\hbar\cE$.}
\end{center}
\end{figure}

Figure \ref{fig:nfparam} shows their results which we have determined 
numerically for the (1,1) and (2,1) bridge-orbit bifurcations for 
$\alpha=2.02$.  In the center panels, we show besides $\tau_0$ also 
the scaled periods $\tau$ of all three periodic orbits.  The parameters 
$a$, $b$, $S_0$ and $\rho$ turn out to be approximately constant 
throughout the bifurcation region, 
as hoped, in particular for the
symmetric (1,1) bridge. 
Note that at $\epsilon=0$, the value
of $b$ is exactly zero for the symmetric bridge bifurcation, so that
the cubic term in \eq{eq:nform3} does not contribute there.
Furthermore, the contribution of the cubic term
$bI^3$ to the actions of the periodic orbits remains much smaller than
that of the quadratic term $aI^2$ throughout the whole bifurcation
region in both cases (1,1) and (2,1). It will therefore be omitted
again in the following.

The combined contribution of all orbits involved in the bifurcation
to the semi\-classical level density is given by \cite{Sieber97,Ozorio87}
\begin{eqnarray}\fl
\delta g(E)&=&\frac{1}{2\pi^2\hbar^2}\Re\int\!\! \rmd q \!\int\!\! \rmd p\, 
  \Psi(q,p,\epsilon)\exp\left[\frac{\rmi}{\hbar}\left\{\hat{S}(q,p,\epsilon)
                              -qp\right\}-\frac{\rmi\pi}{2}\nu\right]\!, 
\label{eq:scld}
\end{eqnarray}
where $\hat{S}(q,p,\epsilon)$ is the generating function defined in
\eq{genfunc} which contains the appro\-priate normal form, and the 
amplitude function $\Psi(q,p,\epsilon)$ is given by
\begin{equation}
\Psi(q,p,\epsilon)=\pp{\hat{S}}{E}
  \left|\pp{^2\hat{S}}{q\partial p}\right|^{1/2}.
\end{equation}
Here $\nu$ is the Maslov index of the bridge orbit labeled by $(n_+,n_-)$. 
It is given by
\begin{equation}
\nu=2n_r\,,\qquad n_r=n_++n_-\,,
\end{equation}
where $n_r$ is equal to the number of librations in the radial direction.
The amplitude function  $\Psi(q,p,\epsilon)$
usually has a moderate dependence on the variables $q,p,\epsilon$ and can 
be replaced by its value at the origin, $\Psi(0,0,0)=\partial S_0/\partial
E=T_0$, which is the average period of the orbit cluster.  By
transforming the variables from $(q,p)$ to $(\varphi,I)$ and 
integrating over $\varphi$, one obtains
\begin{equation}
\delta g(E)\simeq\frac{T_0}{\pi\hbar^2}\Re\, \rme^{\rmi(S_0/\hbar-\pi\nu/2)}
 \int_{-\rho}^{\rho} \rmd I\, \rme^{\rmi(-\epsilon I+aI^2)/\hbar}.
 \label{eq:uniform}
\end{equation}
Here, we omitted the cubic term $bI^3$ in the normal form \eq{eq:nform3}
as stated above.
The integral on the right-hand side can be analytically expressed as
\begin{eqnarray}\fl
\int_{-\rho}^{\rho}\rmd I\,\rme^{\rmi(-\epsilon I+aI^2)/\hbar}
=\sqrt{\frac{2\pi\hbar}{4a}}\,\rme^{\rmi\epsilon^2\!/4a\hbar}
\left\{\sigma_+(c_++\rmi s_+)+\sigma_-(c_-+\rmi s_-)\right\},
\label{eq:uniform2} \\ \fl
c_\pm=\rmC(|x_\pm|)\,, \;\quad s_\pm=\rmS(|x_\pm|)\,, \;\quad
x_\pm=\sqrt{\frac{4a}{2\pi\hbar}}\left(\rho\mp\frac{\epsilon}{2a}\right),
\;\quad \sigma_\pm=\sgn x_\pm\,,
\label{eq:uniform3}
\end{eqnarray}
where $\rmC(x)$ and $\rmS(x)$ are the Fresnel functions defined by
\begin{equation}
\rmS(x)=\int_0^x\sin\left(\frac{\pi}{2}t^2\right)\rmd t\,, \qquad
\rmC(x)=\int_0^x\cos\left(\frac{\pi}{2}t^2\right)\rmd t\,.
\label{eq:fresnelint}
\end{equation}
Inserting \eq{eq:uniform2} into \eq{eq:uniform}, one obtains the 
{\it local uniform approximation} to the level density:
\begin{eqnarray}\fl
\delta g(E)=\frac{T_0}{\hbar^{3/2}\sqrt{2\pi a}}
\Re\, \rme^{\rmi(S_{\rm B}/\hbar-\pi\nu/2)}
\left\{\sigma_+(c_++\rmi s_+)+\sigma_-(c_-+\rmi s_-)\right\}.
\label{eq:localua}
\end{eqnarray}
In terms of the scaled dimensionless normal form parameters 
given in \eq{eq:dlp}, the scaled level density becomes
\begin{eqnarray}\fl
\delta g(\cE)=\sqrt{\frac{\cE}{2\pi\tilde{a}}}\,\tau_0
\Re\, \rme^{\rmi(\tau_{\rm B}\cE-\pi\nu/2)}
\left\{\sigma_+(c_++\rmi s_+)+\sigma_-(c_-+\rmi s_-)\right\},
\end{eqnarray}
where $\tau_{\rm
B}=S_{\rm B}/\hbar\cE=\tau_0+\tilde{\epsilon}^2/4\tilde{a}$ is the
scaled period of the bridge orbit, $c_\pm$ and $s_\pm$ are given by
\eq{eq:uniform3} with the arguments
\begin{equation}
x_\pm=\sqrt{\frac{2\tilde{a}\cE}{\pi}}
  \left(\tilde{\rho}\mp\frac{\epsilon}{2\tilde{a}}\right).
\end{equation}

\subsection{Global uniform approximation}

The local uniform approximation \eq{eq:localua} is valid only
in the vicinities of the bifurcation points, where the properties of
periodic orbits are nicely described by the normal form \eq{eq:nform1}. 
Far from the bifurcation points in the $\epsilon$ variable, the 
standard asymptotic trace formulae should work, which are given in the 
form of the Gutzwiller formula \cite{Gutzwiller} for the isolated 
orbits and the Berry-Tabor formula \cite{BerryTabor} for the family 
of bridge orbits.  The purpose of the so-called {\it global uniform
approximations} is to interpolate between the local result
\eq{eq:localua} and the standard trace formulae.  For the generic 
bifurcations, such global uniform approximations were developed by 
Sieber and Schomerus \cite{Sieber96,Schomerus97a,Sieber98}.  We shall
presently follow their procedure to derive a global uniform
approximation for the integrable bridge bifurcations.

In order to go beyond the local uniform approximation, one can
start from equation \eq{eq:scld}, but one has to expand the
amplitude function $\Psi(q,p,\epsilon)$ around the bifurcation
point in a way similar to the normal form function $S(q,p,\epsilon)$
for the action integral.  We find that it is sufficient here to take 
into account only a linear term in $I$.  We thus write $\delta g(E)$ as
\begin{eqnarray}\fl
\delta g(E)=\frac{1}{\pi\hbar^2}\Re\int_{-\rho}^{\rho}
\rmd I(\alpha+\beta I)\exp\left[\frac{\rmi}{\hbar}
  \left\{\hat{S}(q,p)-qp\right\}-\frac{\rmi\pi}{2}\nu\right].
\label{globuni}
\end{eqnarray}
In principle, the parameters $\alpha$ and $\beta$ are determined from
higher-order expansion coefficients in the normal form
(cf. \cite{Schomerus97a}), but in practice one may determine them from
the condition that the integral \eq{globuni} reproduces the asymptotic
contributions of the bifurcating orbits to the standard trace formulae
far from the bifurcation points.  There, the action differences
between different periodic orbits corresponding to the stationary
points of the normal form \eq{eq:nform1} are much larger than $\hbar$.
This is equivalent to taking the asymptotic expansions of the Fresnel
integrals when their arguments are much larger than unity,
\[
x_\pm=\sqrt{\frac{2(a\rho^2\mp\epsilon\rho+\epsilon^2/4a)}{\pi\hbar}}
=\sqrt{\frac{2}{\pi}\frac{S_\pm-S_{\rm B}}{\hbar}}\gg 1\,.
\]
Their asymptotic forms are (see e.g., \cite{abro})
\begin{eqnarray}
\rmC(x)\simeq \frac12+\frac{1}{\pi x}\sin\frac{\pi}{2}x^2, \qquad
\rmS(x)\simeq \frac12-\frac{1}{\pi x}\cos\frac{\pi}{2}x^2, \NN
\rmC(x)+\rmi\rmS(x)\simeq \frac{\rme^{\rmi\pi/4}}{\sqrt{2}}-\frac{\rmi}{\pi x}
 \rme^{\rmi\pi x^2/2}, \qquad x\gg 1\,. \label{eq:asymptfi}
\end{eqnarray}
Their contribution to the level density is then given by
\begin{eqnarray}\fl
\delta g(E)=\frac{1}{\pi\hbar^2}\Re \rme^{\rmi(S_0/\hbar-\pi\nu/2)}
\int_{-\rho}^{\rho} \rmd I(\alpha+\beta I)
 \rme^{\rmi(-\epsilon I+aI^2)} \NN \fl\phantom{\delta g(E)}
=\frac{1}{\pi\hbar^2}\Re\Biggl[\sqrt{\frac{2\pi\hbar}{4a}}
\left(\alpha+\frac{\beta\epsilon}{2a}\right)
 \rme^{\rmi(S_{\rm B}/\hbar-\pi\nu/2)}
\left\{\sigma_+(c_++\rmi s_+)+\sigma_-(c_-+\rmi s_-)\right\} \NN
+\frac{\beta\hbar}{2a}\left(
 \rme^{\rmi(S_+/\hbar-\pi(\nu+1)/2)}
 -\rme^{\rmi(S_-/\hbar-\pi(\nu+1)/2)}\right)\Biggr],
\label{eq:global1}
\end{eqnarray}
with $S_{\rm B}$ and $S_\pm$ given by \eq{eq:action1}.
Inserting \eq{eq:asymptfi}, one has
\begin{eqnarray}
\delta g(E) &\simeq&
 \frac{\alpha+\frac{\epsilon\beta}{2a}}{\pi\hbar^2\sqrt{\pi a/\hbar}}
 \left(\frac{\sigma_++\sigma_-}{2}\right)
 \cos\left(\frac{S_{\rm B}}{\hbar}-\frac{\pi}{2}\nu
 +\frac{\pi}{4}\right) \NN
 &&+\frac{\alpha+\beta\rho}{\pi\hbar|2a\rho-\epsilon|}
 \cos\left(\frac{S_+}{\hbar}-\frac{\pi}{2}(\nu+\sigma_+)\right) \NN
 &&+\frac{\alpha-\beta\rho}{\pi\hbar|2a\rho+\epsilon|}
 \cos\left(\frac{S_-}{\hbar}-\frac{\pi}{2}(\nu+\sigma_-)\right).
\label{eq:global2}
\end{eqnarray}
The first term on the right-hand side can be identified as the
contribution of the bridge orbit with its Berry-Tabor amplitude
\begin{equation}
A_{\rm BT}=\frac{T_0}{\pi\hbar^2}
  \sqrt{\frac{2\pi\hbar}{|\cK|}},\qquad \cK=\pp{^2S}{I^2},
\end{equation}
and we have
\begin{equation}
\alpha+\frac{\epsilon\beta}{2a}=T_{\rm B}\,, \qquad
a=\frac12\,\cK\,, \label{eq:nfcurv}
\end{equation}
$T_{\rm B}$ being the period of the primitive bridge orbit.
The second and third terms in the right-hand side of
\eq{eq:global2} are identified as the contributions of
two isolated orbits with their Gutzwiller amplitudes
\begin{equation}
A_{\pm}=\frac{T_\pm}{\pi\hbar n_\pm\sqrt{|2-\TrM_\pm|}},
\label{eq:uamplitude}
\end{equation}
where $T_\pm$ are the full periods of these orbits and $n_\pm$ are
their repetition numbers.  Furthermore one finds
\begin{equation}
\alpha\pm\beta\rho=T_\pm\,, \qquad
2a\rho\pm\epsilon=\sigma_\pm n_\pm\sqrt{|2-\TrM_\pm|}\,.
\label{eq:nfparam2}
\end{equation}
From \eq{eq:uamplitude} and \eq{eq:nfparam2}, we can determine the
parameters $\alpha$ and $\beta$ by
\begin{eqnarray}
\alpha =\frac12\,(T_+ + T_-)\,,\qquad
\frac{\beta}{2\pi\hbar a}
=\frac{T_+ - T_-}{\frac{\sigma_+T_+}{A_+}+\frac{\sigma_-T_-}{A_-}}.
\end{eqnarray}
Inserting them into (\ref{eq:global1}), we obtain the {\it global uniform
approximation} for the level density
\begin{eqnarray}\fl
\delta g(E)=\frac{A_{\rm BT}}{\sqrt{2}}
\Re\left[\rme^{\rmi(S_{\rm B}/\hbar-\pi\nu/2)}
\left\{\sigma_+(c_++\rmi s_+)+\sigma_-(c_-+\rmi s_-)\right\}\right] \NN
\fl\phantom{\delta g(E)=}
+\frac{T_+-T_-}{\frac{\sigma_+T_+}{A_+}+\frac{\sigma_-T_-}{A_-}}
\left\{\cos\left(\frac{S_+}{\hbar}-\frac{\pi}{2}(\nu+1)\right)
-\cos\left(\frac{S_-}{\hbar}-\frac{\pi}{2}(\nu+1)\right)\right\}
\label{eq:global3}
\end{eqnarray}
with the arguments of the Fresnel integrals given by
\begin{equation}
x_\pm=\sqrt{\frac{2}{\pi}\frac{S_\pm-S_{\rm B}}{\hbar}}.
\end{equation}
Note that all quantities entering this formula are determined from
the invariant properties of the periodic orbits.

The global uniform approximation \eq{eq:global3} is the main
result of this section.  We shall refer to it in the following
as UA2.  It becomes important in particular when the two bifurcations,
at which the bridge orbits are created and annihilated, are close to 
each other so that neither of the isolated circular orbits are in 
the asymptotic region; the two bifurcations then cannot be treated
separately.  Sufficiently far from the bifurcation points, where all
orbits reach their asymptotic domains, one can use the asymptotic 
trace formulae, referred to as ASY in the following, i.e., the 
Gutzwiller formula \cite{Gutzwiller} for the isolated circular
orbits and the Berry-Tabor formula \cite{BerryTabor} for the bridge 
orbit families.  If this is also the case for a central region
{\it between} the two bifurcations, where the bridges are present,
their bifurcations from/into the isolated orbits can be treated separately 
in the uniform approximation given in \cite{Kaidel04a}, which we
shall refer to as UA1.

\subsection{Numerical results}

In the following, we test numerically the various semiclassical
approximations to the level density.  We restrict ourselves to the
{\it coarse-grained} level density in the scaled energy variable $\cE$, 
which quantum-mechanically is defined in terms of the scaled energy 
spectrum $\{\cE_i\}$ by a sum over normalized Gaussians with width $\gamma$:
\begin{equation}
g_\gamma(\cE) = \frac{1}{\sqrt{\pi}\gamma}\sum_i
  \exp\{-(\cE-\cE_i)^2\!/\gamma^2\}.
\end{equation}
In the semiclassical trace formulae, each contribution of a periodic 
orbit (po) then has to be multiplied by an exponential damping factor
$\exp\{-(\gamma \tau_{\rm po}/2)^2\}$
(see e.g., \cite{BrackText}), so that the contributions of longer
orbits are suppressed and the sum over the periodic orbits converges.  
In our numerical results given below, we have used $\gamma=0.3$ in 
all cases.  With this value of $\gamma$, the condition for the 
above damping factor to be smaller than $10^{-2}$ corresponds to
$\tau_{\rm po}>14$, whose contributions can be safely neglected.

\begin{figure}
\begin{center}
\includegraphics[width=.9\linewidth]{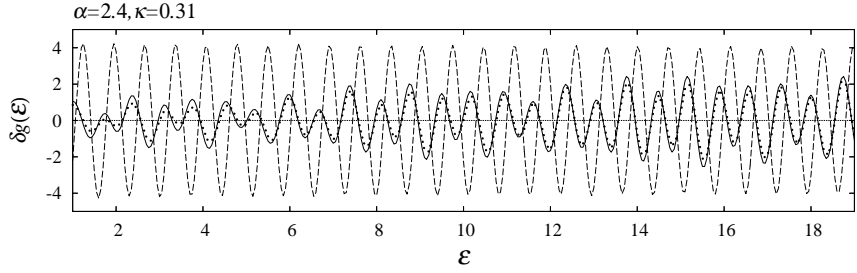}
\end{center}
\caption{\label{fig:asympt}
Oscillating part of the scaled level density $\delta g(\cE)$ for the
Hamiltonian (\ref{eq:hamil1}) with $\alpha=2.4$ and $\kappa=0.31$.
Solid, dashed and dotted curves represent UA1, ASY (divided by 10) and
QM, respectively. In the semiclassical level densities, the periodic 
orbit sum includes the C$_+^{n_+}$ orbits with $n_+\leq 3$, the
orbit C$_-$, and the bridge B(2,1).}
\end{figure}
In figure~\ref{fig:asympt} the oscillating part of the scaled level
densities $\delta g(\cE)$, labeled by ASY and UA1, are compared with
the quantum-mechanical result (QM) for the system \eq{eq:hamil1} with 
$\alpha=2.4$ and $\kappa=0.31$.  These values of the parameters 
correspond to a point close to the left bifurcation of the bridge orbit
B(2,1).  In calculating the semiclassical level densities, we take into 
account the circular periodic orbits C$_+^{n_+}$ with repetition
numbers $n_+\leq 3$, the orbit $C_-$ and the bridge family B(2,1) 
(see figure~\ref{fig:bridge1} for the values of their scaled
periods). 
We see that the uniform approximation UA1 for the bifurcating
circular orbits with the bridge orbit improves the semiclassical level
density over the asymptotic one and nicely reproduces the quantum
results near the bifurcation points.

\begin{figure}
\begin{center}
\includegraphics[width=.9\linewidth]{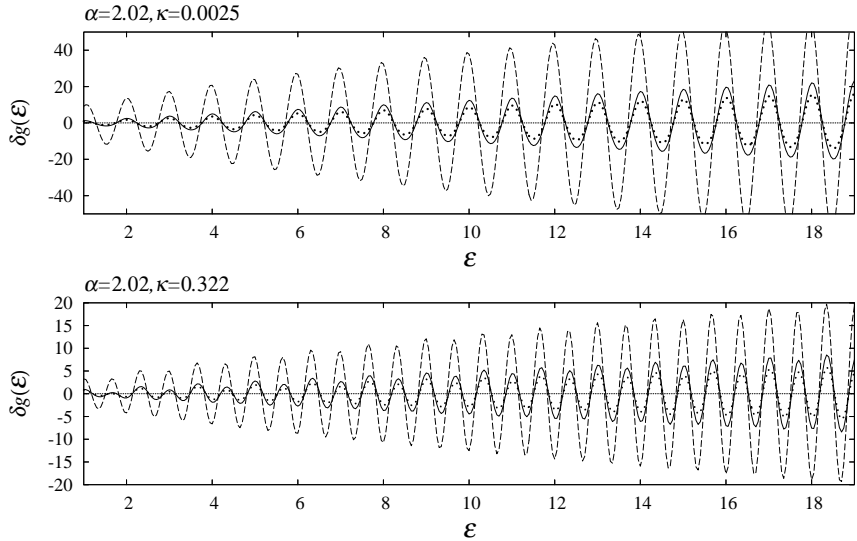}
\end{center}
\caption{\label{fig:uniform1}
Comparison of uniform approximations for $\alpha=2.02$.  Solid, dashed
and dotted curves represent UA2, UA1 and QM, respectively.
Upper and lower panels are calculated at the bifurcation points of 
symmetric bridge (1,1) and asymmetric bridge (2,1), respectively.
In the upper panel, the periodic orbit sum in the
semiclassical level density is taken over C$_\pm^{n_\pm}$ and the
bridge B$(n_+,n_-)$ with $n_\pm\leq3$, while in the lower panel,
C$_+^{n_+}$ with $n_+\leq 3$, C$_-$ and the bridge B(2,1) are taken
into account.}
\end{figure}
Figure~\ref{fig:uniform1} shows the result of the global uniform
approximation UA2 for $\alpha=2.02$.
$\kappa=0.0025$ and $\kappa=0.322$ correspond to the
bifurcation points of the bridge orbits B(1,1) and B(2,1), respectively.
In the periodic orbit sum, all periodic orbits with $\tau<15$ are included;
namely, C$_\pm^n$ and bridge B$(n,n)$ with $n\leq 3$ for $\kappa=0.0025$,
and C$_+^{n_+}$ with $n_+\leq 3$, C$_-$ and B(2,1) for $\kappa=0.322$.
In comparison to the separate
uniform treatment of the two bifurcations (approximation UA1), 
the UA2 formula \eq{eq:global2} reasonably improves the level density, 
but we obtain a slight overestimation.  Let us consider the origin of 
this deviation.  We note that for $\alpha\sim2$, there is no 
asymptotic region between the two bifurcation points.  Therefore, the 
procedure to determine the parameters to reproduce the Berry-Tabor 
asymptotic form is not justified.  In the UA2, the normal form
parameter $a$ is determined such that the asymptotic form reproduces 
the Berry-Tabor trace formula.  This corresponds to using the 
curvature $\cK$ of the torus for the normal form parameter $a$ as
in \eq{eq:nfcurv}.
\begin{figure}
\includegraphics[width=\linewidth]{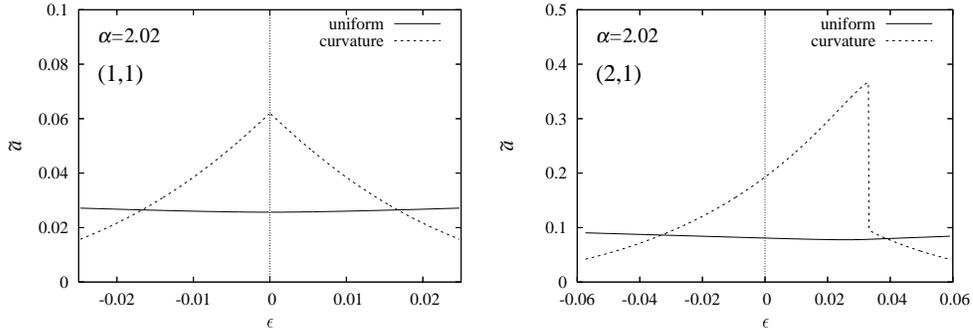}
\caption{\label{fig:nform}
Normal form parameter $\tilde{a}=\hbar\cE a$ (see
(\ref{eq:dlp})) determined from
(\ref{eq:nfparam}) (solid curve) and that from
(\ref{eq:nfcurv}) (dashed curve).  Left and right panels show the
results for (1,1) and (2,1) bridge orbits, respectively.}
\end{figure}
\begin{figure}[p]
\begin{center}
\includegraphics[width=.9\linewidth]{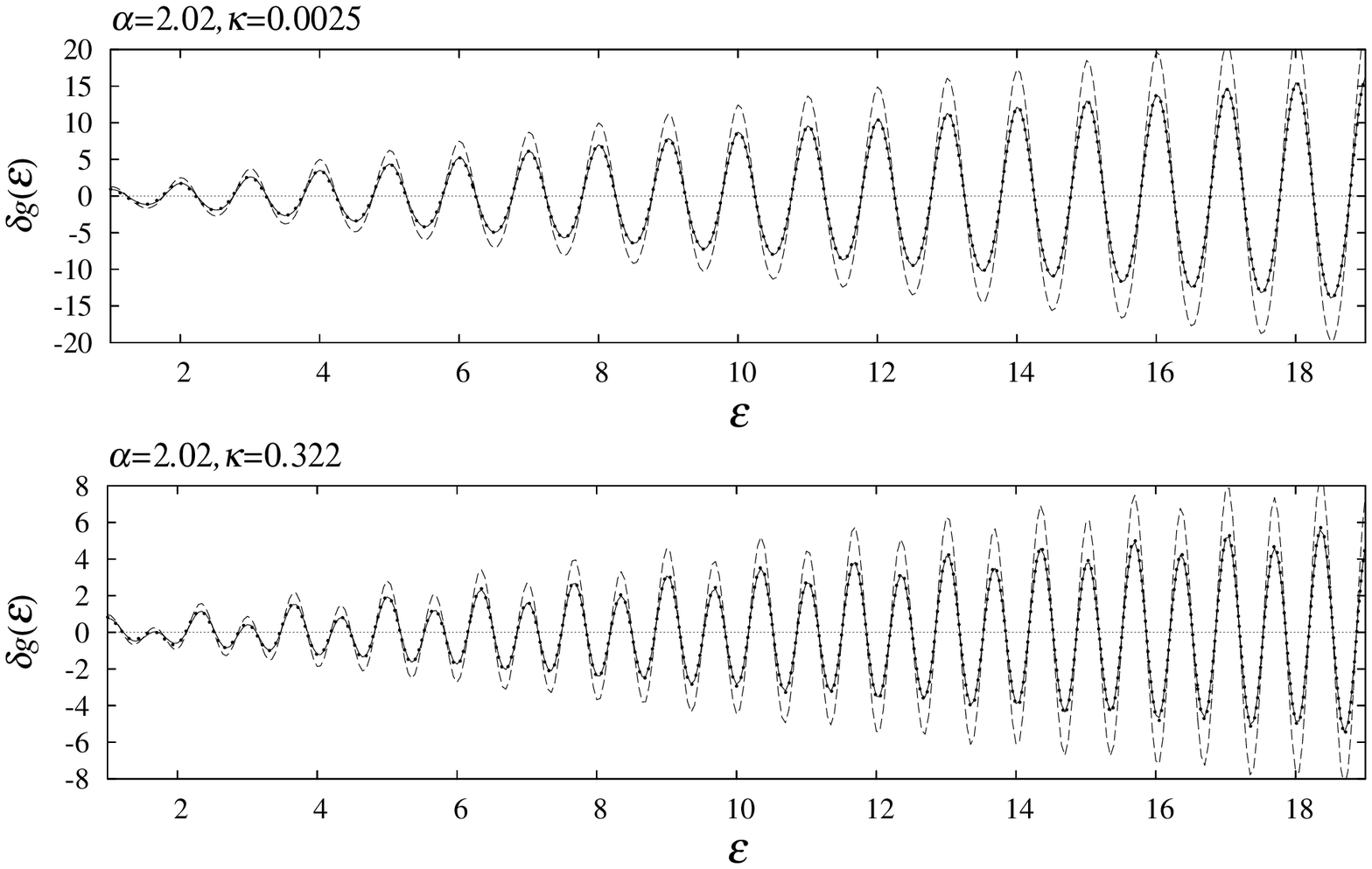}
\end{center}
\caption{\label{fig:uniform2}
Comparison of the global and local uniform approximations.  Solid,
dashed and dotted curves represent uniform approximations UA2L, UA2
and QM, respectively, calculated at the bifurcation
points of (1,1) and (2,1) bridge orbits.
The periodic orbits included in the semiclassical level
densities are same as in figure~\ref{fig:uniform1}.}
\medskip

\begin{center}
\includegraphics[width=.9\linewidth]{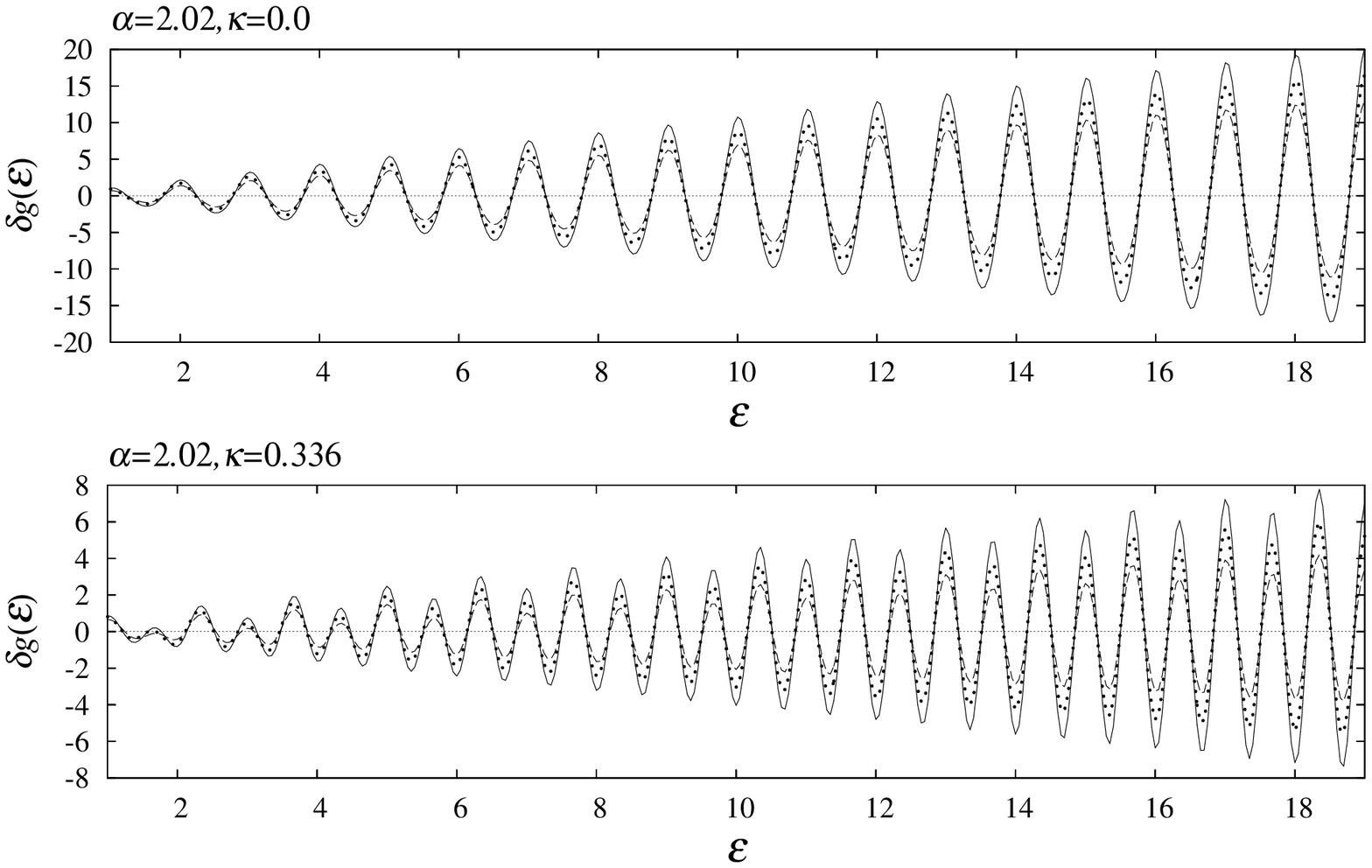}
\end{center}
\caption{\label{fig:uniform3}
Same as figure~\ref{fig:uniform2}, but calculated at the middle of
two bifurcation points of (1,1) and (2,1) bridge orbits.}
\end{figure}
Figure~\ref{fig:nform} compares the values of $a$ determined by
the two methods, i.e., by \eq{eq:nfcurv} and \eq{eq:nfparam}.  They are
significantly different from each other, and the global uniform 
approximation (UA2) does not coincide with the local uniform 
approximation \eq{eq:localua} in the bifurcation region.

Using the local uniform approximation (\ref{eq:localua}) with the normal
form parameters given by (\ref{eq:nfparam}) (referred to as UA2L), the
results are much improved near the bifurcation points as shown in
figure~\ref{fig:uniform2}, but at the middle of the bridge, the quantum
results lie between UA2 and UA2L.

Our numerical results can be interpreted as follows.  The normal form
\eq{eq:nform1} or \eq{eq:nform3} is constructed to reproduce the
bifurcation properties of the three participating periodic orbits.
Therefore, it is most reliable around the bifurcation points
$\epsilon\sim\epsilon_{\rm bif}$.  However, the bridge orbit undergoes
large changes between the two bifurcation points which are globally 
separated in the phase space, and therefore higher-order terms in $I$ 
in the normal form could contribute significantly in the middle region
of the bridge.  In fact, the curvature $\cK$ of the torus, which is
an invariant property of the periodic orbit family and important for 
determining its contribution to the level density, is not correctly 
described by the form \eq{eq:nform3}, as shown in
figure~\ref{fig:nform}.  When the parameter $a$ is shifted toward the 
value determined by the curvature $\cK$, the agreement between the
local uniform approximation and the quantum results becomes better.  
This could be achieved by normal forms with higher-order terms in
$I$ which, however, would render the global uniform approximation
more complicated and less analytic.
 
\section{Summary}
\label{sec:summary}

We have investigated the appearance of bridge orbits, which connect 
two isolated orbits via two successive bifurcations near 
the points where their periods and stabilities coincide,
in various Hamiltonian systems with two degrees of freedom.
In a class of integrable systems, the bridge orbits form
degenerate families.  For these we used a mapping derived from the 
Poincar\'e variables of the isotropic harmonic oscillator to derive
a very simple normal form from which all the invariant properties of
the participating periodic orbits can be derived analytically. 
Using this normal form, we have derived analytical uniform
approximations for the semiclassical level density of the
corresponding quantum systems.  Although the normal form parameters
could not be determined uniquely and their values undergo slight
variations between the two bifurcations, the numerical agreement
between the semiclassical and the quantum-mechanical coarse-grained
level densities is very satisfactory. 

We expect that the remaining differences and the slight variations
of the normal form parameters can be reduced by including more
terms in the normal form.  This would be at the cost of losing the
simple analytical forms of the uniform approximations and of having
to determine more parameters numerically.  

The exploration of suitable normal forms for bridge-orbit bifurcations
in non-integrable systems, such as those shown in figures 
\ref{fig:varsb32} and \ref{fig:erdabridge}, is the subject of further 
studies.  The uniform approximations for non-integrable system form 
also an important subject for understanding the deformed shell structure
in realistic nuclear mean-field models \cite{Arita04,Arita06}.

\ack
K A acknowledges financial support by the Deutsche
Forschungsgemeinschaft (grad\-uate college 638
``Nonlinearity and nonequilibrium in condensed matter'').

\appendix
\section{Derivation of the mapping for bridge orbit bifurcations}
\label{app:mapping}

In this appendix, we derive a mapping from the Poincar\'e variables
$(q,p)$ to action-angle variables $(\varphi,I)$ that are suitable
for the normal forms of bridge orbit bifurcations.
Let us consider the isotropic two-dimensional harmonic oscillator
\begin{equation}
H=\frac12\,(p_x^2+p_y^2)+\frac12\,\omega^2(x^2+y^2)\,.
\end{equation}
All its trajectories are periodic with period $T=2\pi\!/\omega$; they are 
ellipses which may degenerate to a circle or to linear librations. 
We parameterize these periodic orbits using the two constants of 
motion energy $E$ and angular momentum $L=xp_y-yp_x$.
The orbit whose longer semiaxis lies on the $x$ axis is written as
\begin{eqnarray}
x_0(t)=q_c\left(\cos\frac{\theta}{2}+\sin\frac{\theta}{2}\right)
 \cos(\omega t)\,, \\
y_0(t)=q_c\left(\cos\frac{\theta}{2}-\sin\frac{\theta}{2}\right)
 \sin(\omega t)\,, \\
q_c=\sqrt{L_c/\omega}\,, \qquad L_c=E\!/\omega\,,\qquad
L=L_c\cos\theta\,.
\label{orb0}
\end{eqnarray}
Here $L_c$ is the maximum angular momentum at fixed energy, which
is that of the circular orbit running anti-clockwise in the 
$(x,y)$ plane.

Rotating this orbit by an angle $\varphi/2$ about the origin, 
one obtains the most general orbit
\begin{eqnarray}
x(t)=x_0(t)\cos\frac{\varphi}{2}-y_0(t)\sin\frac{\varphi}{2}\,, \NN
y(t)=x_0(t)\sin\frac{\varphi}{2}+y_0(t)\cos\frac{\varphi}{2}\,,\qquad
0 \leq \varphi \leq 2\pi\,.
\end{eqnarray}
The projected Poincar\'e surface of section $\Sigma:=\{ (x(t_i),p_x(t_i))|\;
y(t_i)\!=\!0,\,\dot{y}(t_i)\!\!>\!\!0\}$ at successive times ($i=1,2,\dots$), 
with $t_{i+1}=t_i+T$, defines the Poincar\'e variables
$(q,p)=(x(t_i),p_x(t_i))$.
Each point $(q,p)$ corresponds to an orbit with period $T$, 
and therfore the Poincar\'e variables $(q,p)$ have a one-to-one 
correspondence with the variables $(\theta,\varphi)$, given by
\begin{eqnarray}
q=\frac{q_c\cos\theta}{\sqrt{1-\sin\theta\cos\varphi}}\,, \quad
p=\frac{p_c\sin\theta\sin\varphi}{\sqrt{1-\sin\theta\cos\varphi}}\,,
\quad p_c=\omega q_c\,. \label{eq:pmapping}
\end{eqnarray}
The Jacobian of the transformation $(\varphi,\theta)\to(q,p)$ is
\begin{equation}
\pp{(q,p)}{(\varphi,\theta)}=-\frac12\, p_cq_c\sin\theta\,.
\end{equation}
We now define the action variable $I$ by
\begin{equation}
I=\rho\cos\theta\,, \quad \rho=\frac12\,p_cq_c=\frac12\,L_c\,,
\quad -\rho \leq I \leq \rho\,.
\label{Irho}
\end{equation}
Comparing to \eq{orb0}, one sees that $I=L/2$.  One also sees easily 
that the trans\-formation $(\varphi,I)\to(q,p)$ becomes canonical since
\begin{equation}
\pp{(q,p)}{(\varphi,I)}=1\,.
\end{equation}
For the inverse transformation, one finds from \eq{eq:pmapping} and 
\eq{Irho} the relation
\begin{equation}
I=\frac12\, q\sqrt{4\rho\omega-p^2-\omega^2q^2}\,,\quad
\frac12(p^2+\omega^2 q^2)<2\rho\omega\; (=E)\,.
\end{equation}

\end{document}